\documentclass[letterpaper,twocolumn,10pt]{article}
\usepackage{usenix-2020-09}
\usepackage{tikz}
\usepackage{amsmath}
\usepackage{array}
\usepackage{comment}
\usepackage{filecontents}
\usepackage{tcolorbox}
\usepackage[ruled,vlined,linesnumbered]{algorithm2e}
\usepackage{listings}

\definecolor{codegreen}{rgb}{0,0.6,0}
\definecolor{codegray}{rgb}{0.5,0.5,0.5}
\definecolor{codepurple}{rgb}{0.58,0,0.82}
\definecolor{backcolour}{rgb}{0.95,0.95,0.92}

\let\othelstnumber=\thelstnumber
\def\createlinenumber#1#2{
    \edef\thelstnumber{%
        \unexpanded{%
            \ifnum#1=\value{lstnumber}\relax
              #2%
            \else}%
        \expandafter\unexpanded\expandafter{\thelstnumber\othelstnumber\fi}%
    }
    \ifx\othelstnumber=\relax\else
      \let\othelstnumber\relax
    \fi
}

\lstdefinestyle{customc}{
  belowcaptionskip=1\baselineskip,
  breaklines=true,
  frame=single,
  xleftmargin=0.35cm,
  xrightmargin=0.15cm,
  numbers=left,
  numbersep=5pt,  
  language=C,
  showstringspaces=false,
  basicstyle=\footnotesize\ttfamily,
  keywordstyle=\bfseries\color{green!40!black},
  commentstyle=\itshape\color{purple!40!black},
  identifierstyle=\color{blue},
  stringstyle=\color{orange},
}

\lstdefinestyle{customcArianeExploit1}{
  breaklines=true,
  frame=single,
  xleftmargin=0.4cm,
  xrightmargin=0.2cm,
  numbers=left,
  numbersep=5pt,  
  language=C,
  showstringspaces=false,
  basicstyle=\footnotesize\ttfamily,
  keywordstyle=\bfseries\color{green!40!black},
  commentstyle=\itshape\color{purple!60!black},
  identifierstyle=\color{blue},
  stringstyle=\color{yellow!50!black},
  morekeywords={asm},
  keywordstyle=[2]\bfseries\color{brown!60!black},
}

\lstdefinestyle{customcArianeExploit}{
  breaklines=true,
  frame=single,
  xleftmargin=0.4cm,
  xrightmargin=0.2cm,
  numbers=left,
  numbersep=5pt,  
  language=C,
  showstringspaces=false,
  basicstyle=\footnotesize\ttfamily,
  keywordstyle=\bfseries\color{blue},
  commentstyle=\itshape\color{green!50!black},
  identifierstyle=\color{black},
  stringstyle=\color{brown},
  morekeywords={asm},
  keywordstyle=[2]\bfseries\color{black},
}

\lstdefinestyle{customlog}{
  breaklines=true,
  frame=single,
  xleftmargin=0.35cm,
  xrightmargin=0.15cm,
  numbers=left,
  numbersep=5pt,  
  language=C,
  showstringspaces=false,
  basicstyle=\footnotesize\ttfamily,
  keywordstyle=\color{blue},
  commentstyle=\itshape\color{purple!40!black},
  identifierstyle=\color{blue},
  stringstyle=\color{orange},
  keywords=[2]{INFO},
  keywords=[3]{ERROR},x
  keywordstyle=[2]\bfseries\color{green!40!black},
  keywordstyle=[3]\bfseries\color{red!500!black},
}

\definecolor{verilogcommentcolor}{RGB}{104,180,104}
\definecolor{verilogkeywordcolor}{RGB}{49,49,255}
\definecolor{verilogsystemcolor}{RGB}{128,0,255}
\definecolor{verilognumbercolor}{RGB}{255,143,102}
\definecolor{verilogstringcolor}{RGB}{160,160,160}
\definecolor{verilogdefinecolor}{RGB}{128,64,0}
\definecolor{verilogoperatorcolor}{RGB}{0,0,128}
\definecolor{pointcolor}{RGB}{192,0,0} 
\lstdefinestyle{prettyverilog}{
   language           = Verilog,
   commentstyle       = \color{verilogcommentcolor},
   alsoletter         = \$'0123456789\`,
   literate           = *{+}{{\verilogColorOperator{+}}}{1}%
                         {-}{{\verilogColorOperator{-}}}{1}%
                         {@}{{\verilogColorOperator{@}}}{1}%
                         {;}{{\verilogColorOperator{;}}}{1}%
                         {*}{{\verilogColorOperator{*}}}{1}%
                         {?}{{\verilogColorOperator{? }}}{1}%
                         {:}{{\verilogColorOperator{:}}}{1}%
                         {<}{{\verilogColorOperator{<}}}{1}%
                         {>}{{\verilogColorOperator{> }}}{1}%
                         {!}{{\verilogColorOperator{!}}}{1}%
                         {^}{{\verilogColorOperator{^}}}{1}%
                         {|}{{\verilogColorOperator{|}}}{1}%
                         {||}{{\verilogColorOperator{|| }}}{1}%
                         {=}{{\verilogColorOperator{= }}}{1}%
                         {==}{{\verilogColorOperator{== }}}{1}%
                         {=>}{{\verilogColorOperator{=> }}}{1}%
                         {[}{{\verilogColorOperator{[}}}{1}%
                         {]}{{\verilogColorOperator{]}}}{1}%
                         {(}{{\verilogColorOperator{(}}}{1}%
                         {)}{{\verilogColorOperator{)}}}{1}%
                         {,}{{\verilogColorOperator{,}}}{1}%
                         {.}{{\verilogColorOperator{.}}}{1}%
                         {~}{{\verilogColorOperator{$\sim$}}}{1}%
                         {\%}{{\verilogColorOperator{\%}}}{1}%
                         {\&}{{\verilogColorOperator{\& }}}{1}%
                         {\&\&}{{\verilogColorOperator{\&\& }}}{1}%
                         {\#}{{\verilogColorOperator{\#}}}{1}%
                         {\ /\ }{{\verilogColorOperator{\ /\ }}}{3}%
                         {\ _}{\ \_}{2}%
                        ,
   morestring         = [s][\color{verilogstringcolor}]{"}{"},%
   identifierstyle    = \color{black},
   vlogdefinestyle    = \color{verilogdefinecolor},
   vlogconstantstyle  = \color{verilognumbercolor},
   vlogsystemstyle    = \color{verilogsystemcolor},
   basicstyle         = \scriptsize\fontencoding{T1}\ttfamily,
  columns=fullflexible, 
   keywordstyle       = \bfseries\color{verilogkeywordcolor},
   morekeywords      = {val, when, port, coverage, unique},
   numbers            = left,
   numbersep          = 5pt,
   tabsize            = 2,
   escapeinside       = {/*!}{!*/},
   upquote            = true,
   sensitive          = true,
   showstringspaces   = false, 
   frame              = single,
   breaklines         = true,
   abovecaptionskip   = 0pt,
   belowcaptionskip   = 2pt,   
   xleftmargin        =0.35cm,
   xrightmargin       =0.15cm,
   captionpos         = t,
   emph               = {Point, Point0, Point1, Point2, Point3, Point4, Point5, Point6, Point7, Point8, Point9},
   emphstyle          =\color{pointcolor},
   emph               = {[2] STVEC,SCOUNTEREN,MSTATUS,MTVEC,ML1_ICACHE_MISS,ML1_DCACHE_MISS,MITLB_MISS,MDTLB_MISS,
                             MLOAD,MSTORE,MEXCEPTION,MEXCEPTION_RET,MBRANCH_JUMP,MCALL,MRET,MMIS_PREDICT,MSB_FULL,
                             MIF_EMPTY,MHPM_COUNTER_17,MHPM_COUNTER_18,MHPM_COUNTER_19,MHPM_COUNTER_20,MHPM_COUNTER_21,
                             MHPM_COUNTER_22,MHPM_COUNTER_23,MHPM_COUNTER_24,MHPM_COUNTER_25,MHPM_COUNTER_26,MHPM_COUNTER_27,
                             MHPM_COUNTER_28,MHPM_COUNTER_29,MHPM_COUNTER_30,MHPM_COUNTER_31,property,endproperty, s_eventually}, 
   emphstyle          = {[2]\bfseries\color{verilogkeywordcolor}}
}

\makeatletter

\newcommand\language@verilog{Verilog}
\expandafter\lst@NormedDef\expandafter\languageNormedDefd@verilog%
  \expandafter{\language@verilog}
  
\lst@SaveOutputDef{`'}\quotesngl@verilog
\lst@SaveOutputDef{``}\backtick@verilog
\lst@SaveOutputDef{`\$}\dollar@verilog

\newcommand\getfirstchar@verilog{}
\newcommand\getfirstchar@@verilog{}
\newcommand\firstchar@verilog{}
\def\getfirstchar@verilog#1{\getfirstchar@@verilog#1\relax}
\def\getfirstchar@@verilog#1#2\relax{\def\firstchar@verilog{#1}}

\newcommand\addedToOutput@verilog{}
\lst@AddToHook{Output}{\addedToOutput@verilog}

\newcommand\constantstyle@verilog{}
\lst@Key{vlogconstantstyle}\relax%
   {\def\constantstyle@verilog{#1}}

\newcommand\definestyle@verilog{}
\lst@Key{vlogdefinestyle}\relax%
   {\def\definestyle@verilog{#1}}

\newcommand\systemstyle@verilog{}
\lst@Key{vlogsystemstyle}\relax%
   {\def\systemstyle@verilog{#1}}

\newcount\currentchar@verilog
  
\newcommand\@ddedToOutput@verilog
{%
   \ifnum\lst@mode=\lst@Pmode%
      \expandafter\getfirstchar@verilog\expandafter{\the\lst@token}%
      \expandafter\ifx\firstchar@verilog\backtick@verilog
         \let\lst@thestyle\definestyle@verilog%
      \else
         \expandafter\ifx\firstchar@verilog\dollar@verilog
            \let\lst@thestyle\systemstyle@verilog%
         \else
            \expandafter\ifx\firstchar@verilog\quotesngl@verilog
               \let\lst@thestyle\constantstyle@verilog%
            \else
               \currentchar@verilog=48
               \loop
                  \expandafter\ifnum%
                  \expandafter`\firstchar@verilog=\currentchar@verilog%
                     \let\lst@thestyle\constantstyle@verilog%
                     \let\iterate\relax%
                  \fi
                  \advance\currentchar@verilog by \@ne%
                  \unless\ifnum\currentchar@verilog>57%
               \repeat%
            \fi
         \fi
      \fi
   \fi
}

\lst@AddToHook{PreInit}{%
  \ifx\lst@language\languageNormedDefd@verilog%
    \let\addedToOutput@verilog\@ddedToOutput@verilog%
  \fi
}

\newcommand{\verilogColorOperator}[1]
{%
  \ifnum\lst@mode=\lst@Pmode\relax%
   {\bfseries\textcolor{verilogoperatorcolor}{#1}}%
  \else
    #1%
  \fi
}

\makeatother

\lstdefinestyle{mystyle}{
    commentstyle=\textit,
    keywordstyle=\textbf,
    stringstyle=\color{codepurple},
    basicstyle=\ttfamily,
    breakatwhitespace=false,         
    breaklines=true,      
    frame=single, 
    framexleftmargin=\parindent,
    captionpos=b,                    
    keepspaces=true,                 
    numbers=left,    
    numberstyle=\normalsize,
    stepnumber=1,
    numbersep=5pt,   
    xleftmargin=1.5\parindent,
    showspaces=false,                
    showstringspaces=false,
    showtabs=false,                  
    tabsize=2
}

\usepackage[normalem]{ulem}  

\usepackage{caption}
\usepackage{subcaption}
\usepackage{xurl}
\usepackage{bold-extra}
\usepackage{courier}

\usepackage{bm} 
\usepackage{soul} 

\newcommand{\graph}{\textit{Micro-Event Graph}}
\newcommand{\ourtool}{\textit{WhisperFuzz}}
\newcommand{\hypfuzz}{\textit{HyPFuzz}}

\newcommand{\rc}{{\tt Rocket Core}}
\newcommand{\boom}{{\tt BOOM}}
\newcommand{\cva}{{\tt CVA6}}

\newcommand{\upec}{\textit{UPEC}}
\newcommand{\fadiheh}{\textit{Fadiheh et al.}}
\newcommand{\checkmate}{\textit{Checkmate}}
\newcommand{\sigfuzz}{\textit{SIGFuzz}}
\newcommand{\osiris}{\textit{Osiris}}
\newcommand{\plumber}{\textit{PLUMBER}}
\newcommand{\absynthe}{\textit{ABSynthe}}
\newcommand{\flushreload}{\textsc{Flush+Reload}}
\newcommand{\primeprobe}{\textsc{Prime+Probe}}
\newcommand{\evicttime}{\textsc{Evict+Time}}

\newcommand{\systemverilog}{SystemVerilog}
\newcommand{\verilog}{Verilog}

\newcommand{\vcs}{\textit{VCS}}

\newcommand{\riscv}{RISC-V}

\usepackage{enumitem,amssymb}
\newlist{todolist}{itemize}{2}
\setlist[todolist]{label=$\square$}

\def\todoen{1}
\def\pven{1}
\def\chenen{1}
\def\mren{1}
\def\nken{1}
\def\rken{1}
\def\removesen{0}
\def\adden{0}

\usepackage{graphicx}

\usepackage{diagbox}

\usepackage{multirow}
\usepackage[symbol]{footmisc}

\if 1\todoen 
\newcommand\todo[1]{\textcolor{red}{\textbf{TODO:} #1}}
\else
\newcommand\todo[1]{}
\fi

\if 1\pven
\newcommand\pv[1]{\textcolor{blue}{\textbf{PV:} #1}}
\else
\newcommand\pv[1]{}
\fi

\if 1\chenen
\newcommand\ch[1]{\textcolor{blue}{\textbf{CHEN:} #1}}
\else
\newcommand\ch[1]{}
\fi

\if 1\mren
\newcommand\mr[1]{\textcolor{blue}{\textbf{MR:} #1}}
\else
\newcommand\mr[1]{}
\fi

\if 1\nken
\newcommand\nk[1]{\textcolor{blue}{\textbf{NK:} #1}}
\else
\newcommand\nk[1]{}
\fi

\if 1\rken
\newcommand\rk[1]{\textcolor{blue}{\textbf{RK:} #1}}
\else
\newcommand\rk[1]{}
\fi

\if 1\removesen
\newcommand\red[1]{\textcolor{red!40!white}{\sout{#1}}}
\else
\newcommand\red[1]{}
\fi

\if 1\adden
\newcommand\blue[1]{\textcolor{blue}{#1}}
\else
\newcommand\blue[1]{{#1}}
\fi

\usepackage{tikz}
\newcommand{\tikzxmark}{%
\tikz[scale=0.23] {
    \draw[red,line width=0.7,line cap=round] (0,0) to [bend left=6] (1,1);
    \draw[red,line width=0.7,line cap=round] (0.2,0.95) to [bend right=3] (0.8,0.05);
}}
\newcommand{\tikzcmark}{%
\tikz[scale=0.23] {
    \draw[blue,line width=0.7,line cap=round] (0.25,0) to [bend left=10] (1,1);
    \draw[blue,line width=0.8,line cap=round] (0,0.35) to [bend right=1] (0.23,0);
}}

\usepackage{subcaption} 

\let\othelstnumber=\thelstnumber
\def\createlinenumber#1#2{
    \edef\thelstnumber{%
        \unexpanded{%
            \ifnum#1=\value{lstnumber}\relax
              #2%
            \else}%
        \expandafter\unexpanded\expandafter{\thelstnumber\othelstnumber\fi}%
    }
    \ifx\othelstnumber=\relax\else
      \let\othelstnumber\relax
    \fi
}

\definecolor{lightRed}{RGB}{255, 226, 222}
\newcommand{\Hilight[1]}{\makebox[0pt][l]{\color{lightRed}\rule[-3pt]{#1\linewidth}{10pt}}}

\usepackage[english]{babel}
\newcounter{bug}
\renewcommand*{\thebug}{\textbf{V\arabic{bug}}}

\usepackage[english]{babel}
\newcolumntype{P}[1]{>{\centering\arraybackslash}p{#1}}
\newcolumntype{M}[1]{>{\centering\arraybackslash}m{#1}}

\newcounter{challenge}
\renewcommand*{\thechallenge}{\textbf{C\arabic{challenge}}}
\newenvironment{chal}[1][]{\refstepcounter{challenge}
\noindent\textbf{\thechallenge#1.}\rmfamily}

\newtheorem{defn}{Definition}

\begin{document}
\date{}

\title{\Large WhisperFuzz: White-Box Fuzzing for\\ Detecting and Locating Timing Vulnerabilities in Processors }

\author{
{\rm Pallavi Borkar$^{\S,}$$\thanks{These authors contributed equally to this work.}$, Chen Chen$^{\dagger, \ast}$, Mohamadreza Rostami$^\ddagger$, Nikhilesh Singh$^\S$, Rahul Kande$^\dagger$, }\\
{\rm Ahmad-Reza Sadeghi$^\ddagger$, Chester Rebeiro$^{\S}$, and Jeyavijayan (JV) Rajendran$^\dagger$}\\
$\S$Indian Institute of Technology Madras, India,
$^\dagger$Texas A\&M University, USA, \\
$^\ddagger$Technische Universit\"at Darmstadt, Germany\\
{\tt $\S$\{cs20d202, nik, chester\}@cse.iitm.ac.in,}\\
{\tt $^\dagger$\{chenc, rahulkande, jv.rajendran\}@tamu.edu,}\\
{\tt $^\ddagger$\{mohamadreza.rostami, ahmad.sadeghi\}@trust.tu-darmstadt.de}
} 

\maketitle

\begin{abstract}
    Timing vulnerabilities in processors have emerged as a potent threat. As processors are the foundation of any computing system, identifying these flaws is imperative. Recently fuzzing techniques, traditionally used for detecting software vulnerabilities, have shown promising results for uncovering vulnerabilities in large-scale hardware designs, such as processors. 
    Researchers have adapted black-box or grey-box fuzzing to detect timing vulnerabilities in processors. However, they cannot identify the locations or root causes of these timing vulnerabilities, nor do they provide coverage feedback to enable the designer's confidence in the processor's security.
    
     To address the deficiencies of the existing fuzzers, we present \ourtool{}---the first white-box fuzzer with static analysis ---aiming to detect and locate timing vulnerabilities in processors and evaluate the coverage of microarchitectural timing behaviors. 
     \ourtool{} uses the fundamental nature of processors' timing behaviors, \textit{microarchitectural state transitions}, to localize timing vulnerabilities.
     \ourtool{} automatically extracts microarchitectural state transitions from a processor design at the register-transfer level (RTL) and instruments the design to monitor the state transitions as coverage. Moreover, \ourtool{} measures the time a design-under-test (DUT) takes to process tests, identifying any minor, abnormal variations that may hint at a timing vulnerability. 
   \ourtool{} detects 12 new timing vulnerabilities across advanced open-sourced \riscv{} processors: \boom{}, \rc{}, and \cva{}. Eight of these violate the zero latency requirements of the Zkt extension and are considered serious security vulnerabilities. Moreover, \ourtool{} also pinpoints the locations of the new and the existing vulnerabilities.
\end{abstract}

\section{Introduction}\label{sec:intro}
The evolution in computer architecture has significantly amplified the complexity of hardware design, especially in modern processors, which are the foundation of today's computing systems. 
As technology advances, designers integrate more functionalities into hardware, leading to more intricate architectural and microarchitectural features in processors. 
However, as the complexity of the design increases, so does the number of design regions to verify and protect. 
Traditional techniques to verify modern processors cannot scale with the number of (new) hardware vulnerabilities discovered. 
For example, the number of newly detected hardware common vulnerabilities in the National Vulnerability Database~(NVD) increased from three in 2012 to 92 in 2022~\cite{nvd}.
Further, as of 2023, MITRE reported 117 hardware-related vulnerability types, known as Common Weakness Enumerations~(CWEs)~\cite{MITRE}.  
These rapidly increasing vulnerabilities threaten the security of the expanding digital landscape across different domains necessitating efficient detection strategies\cite{oleksenko2023hide, oleksenko2022revizor, dessouky2019hardfails,chen2022trusting,bloem2022power}.

Timing vulnerabilities are of particular concern as they can leak sensitive information, undermining the entire system's security.
Well-known attacks such as Spectre~\cite{Kocher2018spectre}, Meltdown~\cite{Lipp2018meltdown}, Foreshadow~\cite{van2018foreshadow}, LVI~\cite{vanbulck2020lvi}, RIDL~\cite{ridl}, ZombieLoad~\cite{Schwarz2019ZombieLoad}, CrossTalk~\cite{ragab2021crosstalk}, Zenbleed~\cite{zenbleed}, and Retbleed~\cite{wikner2022retbleed} exploit timing vulnerabilities present in a wide range of commercial processors. Multiple variants of these attacks have been shown to subvert security countermeasures implemented to prevent such attacks.
Unlike functional vulnerabilities, timing vulnerabilities can manifest in a logically correct implementation, making them hard to detect. 
Timing vulnerabilities rely on the difference in execution time of the hardware components to leak sensitive information. 
These vulnerabilities underscore the need for rigorous security analysis in modern processors.
Moreover, unlike software flaws, which can be patched post-deployment, fixing hardware vulnerabilities after manufacturing is difficult, as they are physically ingrained into the Silicon. 
Therefore, detecting vulnerabilities at the pre-Silicon stage is imperative for secure hardware.  

\noindent \textbf{Existing timing vulnerability detection strategies for processors} use formal methods or fuzzing. 
Formal methods, such as theorem proving~\cite{cyrluk1994effective}, model checking~\cite{clarke2018handbook}, assertion proving~\cite{witharana2022survey}, and information-flow tracking~\cite{hu2021hardware} explore design spaces exhaustively and prove security assertions about hardware.
Thus, detecting timing vulnerabilities using formal methods is a rigorous approach to ensure design security~\cite{fadiheh2020formal,trippel2018checkmate}.
However, these methods are limited by the state explosion problem~\cite{clarke2018handbook,baier2008principles,clarke2011model,clarke2001progress};
exhaustively exploring the complex modern hardware is computationally hard~\cite{guo2016scalable,chen2023hypfuzz}. 
Some approaches aim to handle this scalability issue by modeling hardware at the higher abstraction level and approximating its timing behavior~\cite{trippel2018checkmate}.
However, abstracting hardware can lead to over-optimistic results or false positives~\cite{chen2023hypfuzz}.
Furthermore, these formal approaches require a comprehensive understanding of the designs' security specifications and manually defining properties,
an error-prone process~\cite{orenes2021autosva}.

Alternatively, hardware fuzzing has shown its effectiveness in detecting vulnerabilities in large-scale designs~\cite{rfuzz,hur2021difuzzrtl,kandethehuzz,chen2023hypfuzz,xu2023morfuzz, chen2023psofuzz}. 
Using fuzzing, Google detected the recent vulnerability on AMD Zen2 processors, \textit{Zenbleed}~\cite{zenbleed}, a speculative execution vulnerability that allows attackers to extract sensitive information through software exploitation~\cite{googlezenbleed}.
Black-box fuzzing~\cite{weber2021osiris,ibrahim2022microarchitectural} and grey-box fuzzing~\cite{rajapaksha2023sigfuzz} have been applied to detect timing vulnerabilities in processors. 
They explore the design spaces by generating different combinations of instructions as inputs and use performance counters to identify potential timing vulnerabilities~\cite{rajapaksha2023sigfuzz, weber2021osiris, ibrahim2022microarchitectural}. 
While these techniques overcome the scalability issue of formal verification, they suffer from two critical limitations.
First, although they successfully find instructions that cause timing vulnerabilities, they rely on confirmation from designers to identify and pinpoint the root cause (locations)~\cite{weber2021osiris, ibrahim2022microarchitectural, rajapaksha2023sigfuzz}. Second, they lack the adequate coverage metric to capture the timing behaviors of the processor. 
Designers rely on coverage metrics to obtain the necessary confidence before tape-out~\cite{verifiwhitepaper,gopinath2014code,ivankovic2019code}. 
Therefore, introducing such metrics to evaluate the progress of fuzzing is typical~\cite{wile2005comprehensive}. 
We will elaborate on these shortcomings in Section~\ref{sec:related_work}.

\noindent
\textbf{Our Goals and Contributions.}  
We enhance existing fuzzing strategies to address their limitations by integrating static analysis. 
This allows us to automatically pinpoint the sources of timing vulnerabilities and compute the coverage of timing behaviors in processor designs.
Our fuzzer efficiently explores the design space, detecting timing vulnerabilities, while our novel static approach identifies the root causes and provides timing behavior coverage.

Locating the root cause of timing vulnerabilities and computing timing coverage is non-trivial and poses several challenges: 
(i) expressing the timing behaviors of processor modules formally is complex, as they do not operate in isolation and can influence each other; (ii) finer measurement of module timing behaviors is needed, which is time-consuming for modern processors with numerous modules~\cite{cva6, boom}; (iii) tracing vulnerabilities to their root causes within the design space is intricate; and (iv) traditional mutation algorithms used in fuzzers are insufficient for detecting timing vulnerabilities due to their reliance on microarchitectural state transitions.

To address these challenges, (i) We have developed the \graph{}, a static program analysis technique that formally expresses module timing behaviors in a processor by extracting microarchitectural state transitions of a design-under-test (DUT) at the register-transfer level (RTL). To efficiently cover the extensive design space, we tailor the technique to generate individual graphs for each RTL module (cf. Section~\ref{sec:method_graph}). 
(ii) We analyze each RTL module's simulation trace to measure its timing behaviors precisely. To streamline our analysis efforts, we devise a hierarchical strategy based on the characteristics of timing vulnerabilities to prioritize modules for examination (cf. Section~\ref{sec:method_measure_time}).
(iii) We pinpoint the root causes of detected timing vulnerabilities utilizing static analysis techniques and properties of the \graph{}, employing a module-wise strategy to navigate the complex design space.
(iv) We have adapted traditional hardware fuzzing methods to efficiently explore a DUT's design spaces and crafted a specialized mutation engine to exploit timing vulnerabilities. Furthermore, we have instrumented graphs into the DUT to monitor module state transitions based on the input (cf. Section~\ref{sec:method_fuzz}).

\noindent In summary, our contributions are:
\begin{itemize}[align=parleft,leftmargin=*]
    \item We present a novel white-box fuzzer with static analysis, \ourtool{}, for timing vulnerability detection in processors at the RTL. \ourtool{} extracts and monitors microarchitectural state transitions at RTL and measures the timing behaviors of each RTL module to identify timing vulnerabilities. Hardware fuzzing enables \ourtool{} to explore the microarchitectural state space efficiently.
    \item With static analysis, \ourtool{} will identify the locations/root causes of timing vulnerabilities. Moreover, \ourtool{} introduces a timing coverage metric to help designers evaluate the timing behaviors explored.
    \item We evaluate the effectiveness of \ourtool{} on three real-world, open-sourced processors from RISC-V instruction set architecture (ISA) -- \boom{}~\cite{boom}, \rc{}~\cite{rocket_chip_generator}, and \cva{}~\cite{cva6}, which are widely used as benchmarks in the hardware security community.
    \item \ourtool{} finds 12 new timing vulnerabilities across all three benchmarks. Eight of them pose serious security vulnerabilities, according to the \riscv{} Zkt contract~\cite{riscvzkt}. \ourtool{} also pinpoints the locations of all existing and new vulnerabilities.
\end{itemize}

\section{Background}\label{sec:background}

In this section, we provide a succinct background on hardware fuzzing and microarchitectural timing side channels, which form the basis of \ourtool{}.

\subsection{Hardware Fuzzing}\label{sec:hwfuzzing}

Hardware fuzzing is a dynamic verification technique that iteratively generates testing inputs called tests to verify target hardware~\cite{rfuzz,hur2021difuzzrtl,chen2023hypfuzz}. 
A coverage-feedback fuzzer starts by generating an initial set of tests, called \textbf{seeds}, randomly using a \textit{seed generator}.
When fuzzing processors, these seeds are executable programs with a sequence of instructions~\cite{kandethehuzz, chen2023hypfuzz}. 
The fuzzer simulates the target hardware with these tests using open-source or commercial hardware simulation tools such as Verilator~\cite{verilator} and Synopsys VCS~\cite{vcs}. 

During simulation, the fuzzer collects \textbf{coverage} information that quantifies the activities caused by the test in the hardware. For instance, the coverage information can be represented as transitions of finite-state machines~(FSMs)~\cite{kandethehuzz}. 
Fuzzers either instrument the hardware to add activity monitors~\cite{rfuzz, hur2021difuzzrtl}, or use the existing coverage monitors of the simulation tools~\cite{kandethehuzz, chen2023hypfuzz} to collect this coverage information.
Next, it generates new tests automatically by performing bit manipulation operations, called \textbf{mutations} on the current tests, increasing coverage. The fuzzer iterates over this cycle of test generation and simulation to verify target hardware till desired coverage is achieved.


Fuzzers use a \textit{vulnerability detector} to detect vulnerabilities using differential testing or hardware assertions.
In differential testing, the fuzzer runs a golden reference model along with the target hardware and compares its outputs to detect vulnerabilities~\cite{kandethehuzz,hur2021difuzzrtl}.
Alternatively, some fuzzers insert hardware assertions, i.e., rules describing the expected behavior, in the target hardware and detect vulnerabilities as violations of these assertions~\cite{muduli2020hyperfuzzing}. While existing black-box~\cite{weber2021osiris,ibrahim2022microarchitectural} and grey-box~\cite{rajapaksha2023sigfuzz} fuzzers can detect timing vulnerabilities in processors, these approaches fail to locate the root causes of these vulnerabilities. Further, these techniques can not evaluate the explored timing behaviors. \red{ These limitations delay the mitigation process and leave out possibilities for undiscovered vulnerabilities hampering the designer's confidence. } These limitations delay the mitigation process and hamper the designer's confidence due to absence of a coverage metric.

\subsection{Timing Side-Channel Attacks}\label{sec:tsc}
Timing side-channel attacks exploit measurable variations in the execution time of instructions to glean secret information from victim applications. These timing variations arise from interactions of different operands with the micro-architecture. 
For instance,~\cite{bernstein2005cache} proposed a timing attack to recover AES keys by measuring the cache accesses during encryption across different keys.

\begin{lstlisting}[language=C, label = {listing:inst_seq}, caption={A pair of instruction sequences that have identical instructions but operate on different data. The instruction sequence is a timing vulnerability if the execution times differ.},style=customcArianeExploit,float,belowskip=0pt,aboveskip=0pt,firstnumber=1]
LI   t6, 0x81321        LI   t6, 0x11235    
LI   a7, 0xFEEEE        LI   a7, 0xFFFFF 
LD   t4, 0(a7)          LD   t4, 0(a7)
ADDI a9, a5, t4         ADDI a9, a5, t4
LD   t5, 0(t6)          LD   t5, 0(t6)
\end{lstlisting}

Over the years, multiple techniques to perform timing attacks have developed such as \primeprobe{}~\cite{osvik2006cache}, \evicttime{}~\cite{osvik2006cache}, and \flushreload{}~\cite{gruss2016flush+}. The key idea of all these attacks is to target a specific shared resource, such as the cache memory, and exploit its data-dependent timing behavior. Such attacks consist of a sequence of instructions which when executed with different data take different execution times. For example, consider the pair of instruction sequences in Listing~\ref{listing:inst_seq}. The two sequences are identical but differ in the data they operate upon. The sequence can be considered a timing vulnerability if the execution of the two sequences results in different execution times.

The objective of our work is to develop \ourtool{}, which identifies such instruction sequences and pairs of input data that result in different execution times. Furthermore, \ourtool{} localizes the root cause for the differing execution times, thereby assisting in mitigation.

\section{Methodology}\label{sec:method}
In this section, we first explain the relationship between timing behaviors and transitions of a digital circuit. We introduce the \graph{} that helps capture the microarchitectural transitions at a fine granularity. We use this graph to identify the location of timing vulnerabilities and monitor the timing behaviors covered. Further, we  discuss the challenges of extracting the \graph{}  from a processor design. Finally, we give an overview of our solutions to these challenges and elaborate on each solution.

\subsection{Microarchitectural Transitions and Timing Behaviors }
\label{sec:relation_state_time}

A finite-state machine (FSM) can model a sequential digital circuit. Given an input, the circuit transitions through various states in the FSM to produce the output. Assuming each state in the FSM takes constant time, the sequence of state transitions to generate the output determines the execution time of the circuit for a given input. Thus, we claim: \textit{two different inputs resulting in the same state transition sequences will take the same execution time.} 

\begin{figure}[h]
    \centering
    \includegraphics[width=0.75\columnwidth,trim = 2 2 0 0, clip]{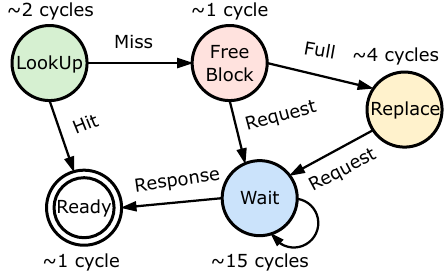}
    \caption{A finite-state machine (FSM) representation of the cache set protocol. Each state is assumed to take a constant time as shown at each node.}
    \label{fig:case_study}
\end{figure}

Consider the case study of a cache composed of multiple cache sets. Each cache set can be represented as an FSM with five states: \texttt{LookUp}, \texttt{FreeBlock}, \texttt{Replace}, \texttt{Wait}, and \texttt{Ready}, as shown in Figure~\ref{fig:case_study}. When a program accesses data, the cache first performs a look-up for the associated memory address in the \texttt{LookUp} state. If the data is found in the cache (cache hit), it will be directly transferred to the \texttt{Ready} state, and the processor can access the data without going to the main memory. 
Thus, an input address that is already present in the cache set causes transitions \texttt{\{LookUp $\rightarrow$ Ready\}} taking three clock cycles in its corresponding implementation. 
However, if the data is not found (cache miss), the cache set will transfer to the \texttt{FreeBlock} state to look for a free cache block to store the data. If a free block exists, the cache will transition to the \texttt{Wait} state and request the memory for the corresponding data. 
In case the cache does not have a free block, it will transfer to the \texttt{Replace} state, select a block for eviction based on the replacement policy, and then transfer to the \texttt{Wait} state. 
The cache waits in this state until it receives data from memory. It then transitions to the \texttt{Ready} state. Thus, if an input address is not present in the cache set, two FSM state transition sequences (and execution times) are possible: (i)~\texttt{\{LookUp $\rightarrow$ FreeBlock $\rightarrow$ Wait $\rightarrow$ Ready\}}, taking 19 clock cycles if the cache has a free block or (ii)~\texttt{\{LookUp $\rightarrow$ FreeBlock $\rightarrow$ Replace $\rightarrow$ Wait $\rightarrow$ Ready\}}, taking 23 cycles if the cache does not have a free block. {Assuming each state in the FSM takes a constant execution time, a difference in the execution time of two inputs implies a difference in the sequence of state transitions followed in the FSM.}
However, in practice, FSM states do not always take a constant execution time. Moreover, an FSM model is abstract and cannot effectively represent complex microarchitectural details in digital circuits. Thus, an FSM representation fails to uncover timing differences 
arising at the microarchitectural level. 
Further, it makes localizing the source of timing difference difficult, delaying the mitigation process.

We introduce \textit{Micro-Event Graphs}~(MEGs) to overcome these drawbacks. A MEG models a given digital circuit using the register-transfer level (RTL) as a set of events, which we call microarchitectural events. Each microarchitectural event affects the contents of at least one element, such as a wire or a register in the digital circuit. A MEG models a given digital circuit as possible events and dependencies between these events. Each node in the MEG represents an element while a directed edge from a parent node to a child node indicates that an event on the parent element can potentially trigger an event on the child element.

\begin{figure}
    \centering
    \includegraphics[width=0.85\columnwidth,trim = 0 0 0 0, clip]{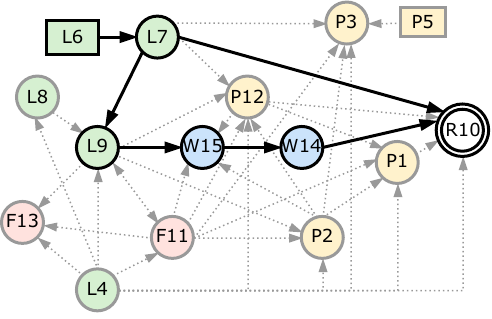}
    \caption{A high level representation of \graph{}~(MEG) for cache set protocol represented as an FSM in Figure~\ref{fig:case_study}.}
    \label{fig:full_graph}
\end{figure}

Figure~\ref{fig:full_graph} shows a high-level representation of the MEG corresponding to the cache set protocol. The MEG consists of 15 nodes and 115 edges. Each state in the corresponding FSM (refer Figure~\ref{fig:case_study}) maps to one or more nodes in the MEG. For example, nodes colored in green (node labels starting with L) correspond to state \texttt{LookUp} and those colored in yellow (node labels starting with P) correspond to state \texttt{Replace}. Each execution of the cache set protocol can be mapped to at least one path in the MEG shown. 
Similar to the state transitions in the FSM, but at the finest granularity, an input triggers a sequence of microarchitectural events in the MEG during its execution, each taking a constant time. 
Thus, one can have the following observations:
\begin{enumerate}
    \item[{\bf P1.}] If two inputs to the microarchitecture result in the same event transitions in the MEG, then the execution time for the two inputs is the same.
    \item[{\bf P2.}] If there is a difference in the execution time of the two inputs, then the sequence of microarchitectural events followed is different for the two executions.
    \item[{\bf P3.}] If two inputs to the microarchitecture result in different event transitions in the MEG, then the execution time for the two inputs may differ.
\end{enumerate}

\subsection{Detecting and Localizing Timing Vulnerabilities in Processors: A High-Level Overview}
The goal of \ourtool{} is to detect and localize timing vulnerabilities in a processor design-under-test (DUT). To detect timing vulnerabilities, we leverage the strength of hardware fuzzing. The fuzzer generates an instruction sequence and at least two corresponding data inputs of the form shown in Listing~\ref{listing:inst_seq}, that take different execution times when applied to the sequence. 
For each pair of instruction sequences and inputs, we trace the path of events in the MEG. For example, in Figure~\ref{fig:full_graph}, each path represents an execution corresponding to different addresses given to a \texttt{load} instruction. These paths are then used to localize the root cause of the vulnerability. The root cause is the event prior to the first bifurcation in these two paths.

\begin{figure*}[t]
    \centering
    \includegraphics[trim=0 640 440 0,clip,width=0.9\linewidth]{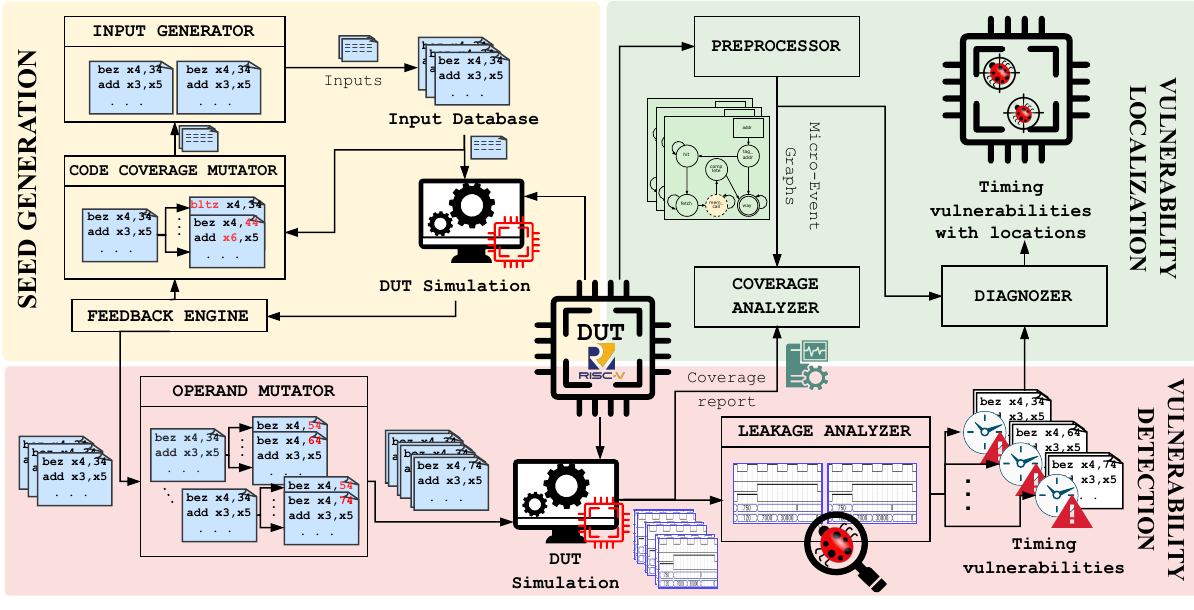}
    \caption{The \ourtool{} framework. It includes three key modules. First, the Seed Generation module internally utilizes a coverage-feedback fuzzer to explore the design space. The generated inputs are recorded in a database. Mutations are performed to improve code coverage. Second, the Vulnerability Detection module uses the generated seed, mutates the instruction operands, and identifies the timing vulnerabilities based on DUT simulations. Finally, the Vulnerability Localization module pinpoints the locations of uncovered vulnerabilities.  }
    \label{fig:frame}
\end{figure*}

For example, consider the two highlighted paths in Figure~\ref{fig:full_graph}. One path traces event sequence \{ L6, L7, R10\}, while the other traces \{L6, L7, L9, W15, W14, R10\}.
Each path corresponds to a different address given to a \texttt{load} instruction. Since the two paths are different, they may take different execution times due to {\bf P3} (See Section~\ref{sec:relation_state_time}), resulting in a timing vulnerability. These paths trace the same events until L7, after which they bifurcate. Hence, an event on node L7 will likely be the vulnerability's root cause. The number of paths covered in the MEG gives a notion of the coverage of timing behaviors of the DUT.

\subsection{Challenges}\label{sec:challenges}
Developing \ourtool{} involves the following challenges.\\

\begin{chal}
[\label{c1}] \textbf{Generating the MEG.}
The MEG must capture all possible events and event transitions in a given processor DUT. This is specifically challenging in modern microprocessors due to their complexity and large code bases. We develop an automated strategy to address this challenge that extracts the MEG given the DUT's source code in RTL form. Section~\ref{sec:method_graph} elaborates on this strategy.
\end{chal}

\begin{chal}[\label{c2}] \textbf{Characterizing timing behavior of each processor module.}
One way to determine timing behaviors is to input instruction sequences to the DUT and measure the execution time. However, a complete processor design contains thousands of signals, complicating the localization of the vulnerability. An alternate bottom-up approach is to isolate each module in the processor, provide inputs, and measure the timing behavior of the module. However, timing differences detected may not be observable when the module is integrated into the complete processor. Therefore, we follow a two-pronged approach where we first generate instruction sequences for the complete processor to detect timing differences and localize at a module level. We present a \textit{hierarchical analysis strategy} to prioritize the modules to be analyzed, thereby detecting vulnerabilities faster, as discussed in Section~\ref{sec:method_measure_time}.

\end{chal}

\begin{chal}[\label{c3}] \textbf{Localizing the source of timing vulnerability.}
The large code space of the processor's DUT makes it challenging to locate the source of a timing vulnerability. It necessitates manual effort and a detailed understanding of the processor's microarchitecture. To address this challenge, we introduce an automated static analysis strategy on the MEG that can identify the root cause of the timing vulnerability within a few seconds. Section~\ref{sec:diagnozer} elaborates on the strategy.
\end{chal}

\begin{chal}[\label{c4}] \textbf{Fuzzing the microarchitectural state space and determining coverage.}
Hardware fuzzing has shown its effectiveness in detecting vulnerabilities in large-scale designs, such as processors. 
However, the existing grey-box processor fuzzers~\cite{rajapaksha2023sigfuzz} are not compatible with timing vulnerability detection. 
Their mutation algorithms are designed to accelerate coverage increment, but timing vulnerability detection requires sequences of data-dependent instructions that change processors' timing behaviors as seen in Listing~\ref{listing:inst_seq}.
Moreover, existing fuzzers lack coverage feedback to demonstrate the timing behaviors covered. Providing such metrics can help designers decide to tape out.
To address this challenge, we develop a specialized mutation algorithm that generates data-dependent instructions.
We then use paths of each module's MEG as the timing coverage metric and instrument the DUT to provide coverage feedback.
\end{chal}

\subsection{The \ourtool{} Framework}

To address the challenges discussed in Section~\ref{sec:challenges}, we develop
\ourtool{} that comprises of three major modules: \textit{Seed Generation}, \textit{Vulnerability Detection}, and \textit{Vulnerability Localization}, as shown in Figure~\ref{fig:frame}.

{\flushleft \bf Seed Generation.} \ourtool{} first uses the Seed Generation module that contains a coverage-feedback fuzzer~\cite{rfuzz,hur2021difuzzrtl,kandethehuzz,chen2023hypfuzz,chen2023psofuzz} to explore design spaces guided by the fuzzer's internal code-coverage metric. The fuzzer utilizes the \textit{Input Generator} to generate input instructions, which are then simulated in the fuzzer's internal \textit{DUT Simulation} unit. The \textit{Feedback Engine} calculates the code-coverage metric of this input. Based on this metric, the \textit{Code Coverage Mutator} mutates the input instructions to improve the coverage. The \textit{Input Database} records various fuzzer-generated inputs and the corresponding code-space covered. 

{\flushleft \bf Vulnerability Detection.} Based on the \textit{Feedback Engine}, \ourtool{} identifies a sequence of instructions that explore new design spaces as seeds and sends them to the Vulnerability Detection module. 
The  \textit{Operands mutator} in the Vulnerability Detection module mutates the data in the instruction sequence to trigger varying timing behaviors~(See Section~\ref{sec:method_fuzz}). These mutated instruction sequences are simulated in the \textit{DUT Simulation} unit of the Vulnerability Detection module.
The \textit{Leakage analyzer} then compares simulation traces and detects timing vulnerabilities (See Section~\ref{sec:method_measure_time}).

{\flushleft \bf Vulnerability Localization.}
\ourtool{} invokes the \textit{Preprocessor} in Vulnerability Localization to extract \graph{} of modules (See Section~\ref{sec:method_graph}) and instruments potential timing behaviors based on the \graph{}. 
For the timing vulnerabilities identified by the \textit{Leakage analyzer}, \ourtool{} invokes the \textit{Diagnozer} to pinpoint the cause of these timing vulnerabilities (See Section~\ref{sec:diagnozer}). The \textit{Coverage analyzer} collects simulation traces of all inputs from the Vulnerability Detection module and calculates the timing behaviors covered. 

The \ourtool{} framework repeats these steps until there is a timeout, or the fuzzer and \textit{Operand mutator} completely cover the timing behaviors of the DUT.

\subsection{Extracting Micro State Transitions}\label{sec:method_graph}

In this section, we elaborate on the MEG used to address challenge~\ref{c1}. Using a \textit{Preprocessor}, \ourtool{} parses the RTL code of a given design to generate its MEG representation. Each node in the MEG represents an event corresponding to an element in the RTL, while an edge indicates that there exists an event on the parent node that can trigger a change in the value of the child node. Edges between nodes are also annotated with conditions required to trigger the event and the RTL line number which causes the dependency between the event. Nodes in the MEG can be categorized into sequential nodes or combinational nodes corresponding to the sequential or combinational nature of the hardware element it represents~\cite{IEEEstd}. Formally, we define a MEG as:

\begin{defn}
Given the RTL of a module $D$ in a DUT, the corresponding micro-event graph is defined as $\mathbb G(D) = ( S, \Sigma)$, where the nodes are denoted by $ S=\{s_1,s_2,\cdots ,s_n\}= \{ S_q \cup   S_c \cup  S_I \cup I \cup O\}$ and include all elements in $D$; $  S_q$ and $  S_c$ represent the sequential and combinational elements respectively; $S_I$ represent external modules instantiated in $D$; $I$ and $O$ are the set of inputs and outputs respectively to the module $D$. The set $\Sigma$ represents the edges in the graph, $\mathbb G$, such that $(s_i,s_j) \in \Sigma$ \red{if and only }if a change in $s_i$ may trigger a change in $s_j$ as per the RTL. Each edge is annotated with the condition required for the change to occur.
\end{defn}

\begin{lstlisting}[language=Verilog, label = {listing:case_study}, caption={Simplified Verilog code of cache set protocol (refer Figure~\ref{fig:case_study}).},style=prettyverilog,float,belowskip=0pt,aboveskip=0pt,firstnumber=1,linewidth=\linewidth, xleftmargin=10pt]
input addr; output way;
reg hit, full, valid, fetch;
wire tag_addr, complete;
assign tag_addr = addr[tag_bits];
always@(posedge clock) begin
    if tag_addr == tag
        way <= index; hit <= 1;
    if hit == 0 && full != 1 && valid == 1 begin
        fetch <= 1; temp_way <= index;
    if fetch == 1: 
        // Call mem_call for fetching block from next level.
        // Sub-instance sets complete signal
        mem_call(addr, complete);
    if complete:
        way <= temp_way;
end

\end{lstlisting}

\begin{figure}[!h]
    \centering
    \includegraphics[width=0.75\columnwidth,trim=0 3 7 0, clip]{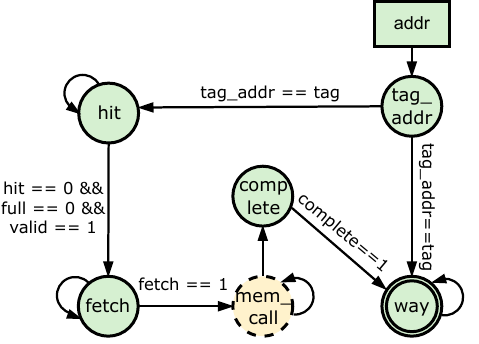}
    \caption{Sub-graph extracted from \graph{} in Figure~\ref{fig:full_graph} of the cache set protocol case study. The L6, L7, L9, W15, W14 and R10 nodes from Figure~\ref{fig:full_graph} correspond to \texttt{addr}, \texttt{tag\_addr}, \texttt{hit}, \texttt{mem\_call}, \texttt{complete}, \texttt{way} respectively.}
    \label{fig:cs_graph}
\end{figure}

Consider the cache set protocol example given in Section~\ref{sec:relation_state_time}, with its corresponding simplified \verilog{} code as in Listing~\ref{listing:case_study}. 
A sub-graph of its MEG is shown in Figure~\ref{fig:cs_graph}, denoting input node \texttt{addr} with an incoming edge. 
Due to the \texttt{assign} statement on Line 4, a transition in the value of \texttt{addr} triggers a change in the value of \texttt{tag\_addr}. 
To denote this dependency, the graph contains a directed edge from the node \texttt{addr} to \texttt{tag\_addr}. 
Similarly, the value of \texttt{tag\_address} can influence the value of \texttt{way} and \texttt{hit} if the condition (\texttt{tag\_address == tag}) on Line 6 holds. 
Hence, the edges from  \texttt{tag\_address} to \texttt{way} and \texttt{hit} are annotated with the condition. The sub-instance call on Line 13 corresponds to the dashed node \texttt{mem\_call} in the MEG. The accept state from node \texttt{way} indicates that it references an output signal. Appendix~\ref{apd:imp_details} provides a further discussion on the implementation details of MEG.

\begin{algorithm}[t]
\SetFuncSty{textsc}
\SetKwInOut{KwIn}{Input}
\SetKwInOut{KwOut}{Output}
\SetKwFunction{getInOutputs}{getInOutputs}
\SetKwFunction{getSignals}{getSignals}
\SetKwFunction{getDestination}{getDestination}
\SetKwFunction{getOperands}{getOperands}
\DontPrintSemicolon
\caption{Preprocessor in Vulnerability Localization module of \ourtool{}.}\label{alg:preproc}
\KwIn{
    $D$ \tcp{RTL code of module}
}
\KwOut{
    $\mathbb G$ \tcp{MEG of module}
}

$\mathbb G(D).I \gets \phi$; $\mathbb G(D).S_I \gets \phi$; $\mathbb G(D).O \gets \phi$\; 
$\mathbb G(D).S \gets \phi$; $\mathbb G(D).\Sigma \gets \phi$\;

$\mathbb G(D).I,~\mathbb G(D).S_I,~\mathbb G(D).O \gets $ \getInOutputs{$D$}\;
$\mathbb G(D).S \gets $ \getSignals{$D$}\;
\tcc{Iterate RTL code}
\For{\textbf{line} $\in D$}{
    \tcp{Get an operand set that can change the value of destination signal}
    $s_2 \gets$ \getDestination{$line$}\; 
    $s_1 \gets$ \getOperands{$line$}\;
    \For{$s_1 \in S_1$}{
        \tcp{Add an edge between each operand and the destination signal}
        $\mathbb G(D).\Sigma \gets \mathbb G(D).\Sigma \cup \{s_1, s_2, line\_no\}$\;   
    }
}

return $\mathbb G$\;

\end{algorithm}

The \textit{Preprocessor} parses the RTL description of the design, translating it into the corresponding graphical representation using Algorithm~\ref{alg:preproc}.  Given an RTL module $D$, the \textit{Preprocessor} first extracts into $\mathbb G(D). X$ where $X = I, O, S, \Sigma, \text{or } S_I$ corresponds to the set of inputs, outputs, nodes, edges, and sub-instance nodes respectively (Lines 3 and 4). 
Finally, the \textit{Preprocessor} iterates over each line in the RTL, identifies destination signals (Line 6) and operand signals (Line 7), and adds an edge between each operand signal and the destination signal (Line 9). The \textit{Diagnozer} ( Section~\ref{sec:diagnozer}) and \textit{Coverage Analyzer} ( Section~\ref{sec:method_fuzz}) of \ourtool{} utilize this MEG for further analysis. \ourtool{} generates \graph{} for each module individually to counteract the vast design complexity of processors. The comparitively reduced permodule complexity consumes a reasonable computation cost and resource utilization as shown in Table~\ref{tab:overhead_stats}.

{\flushleft \bf Micro-Event Path.} We define a \textit{Micro-Event Path}~(MEP) as a path that starts at the input node and traces connected nodes until it reaches the output node. Given an input, this path traces the sequence of events triggered in the module. Formally,

\begin{defn}
A sequence of directed edges, $P = \left \langle (s_{1},s_{2}), (s_{2}, s_{3}), \cdots,(s_{n-1}, s_{n}) \right \rangle$ where  $(s_{i}, s_{j}) \in \Sigma, \forall(i,j)$ and $s_{1} \in I$ and $s_{n} \in O$, is a Micro-Event Path in $\mathbb G(D)$.    
\end{defn}
 
Figure~\ref{fig:cs_graph} notes two different MEPs from the input node \texttt{addr} to the output node \texttt{way}. 
Each path can be mapped to a path in the FSM in Figure~\ref{fig:case_study}, but at a finer granularity. In case of a cache hit, the module follows the FSM path \texttt{\{LookUp $\rightarrow$ Ready\}} mapped to MEP \{\texttt{addr $\rightarrow$ tag\_addr $\rightarrow$ way}\}. In case of a cache miss, if the cache set has a free cache line, the module follows the FSM path \{\texttt{LookUp $\rightarrow$ FreeBlock $\rightarrow$ Wait $\rightarrow$ Ready}\} mapped to \{\texttt{addr $\rightarrow$ tag\_addr $\rightarrow$ hit $\rightarrow$ fetch $\rightarrow$ mem\_call $\rightarrow$ complete $\rightarrow$ way}\}. A comparison of these two differing MEPs indicates the RTL wire \texttt{tag\_addr} as the hardware element responsible for the divergence in the two paths. \ourtool{} utilizes this information to trace the root cause of the vulnerability, as elaborated in Section~\ref{sec:diagnozer}.

\subsection{Characterizing Timing Behaviors}\label{sec:method_measure_time}
Existing hardware fuzzers detect timing vulnerabilities by measuring the execution time of an entire processor through performance counters~\cite{weber2021osiris, ibrahim2022microarchitectural, rajapaksha2023sigfuzz, weaver2013non}. 
\ourtool{}, however, requires finer information to  localize the timing vulnerabilities. 
Therefore, \ourtool{} computes the execution time taken by each module within the processor. For this purpose, we use simulation of inputs generated by the {\em Seed Generation} and mutated by the {\em Operand Mutator}. 

For each simulation of the DUT, the \textit{Simulator} generates a set of simulation traces corresponding to each signal within the DUT. A simulation trace is a time series that records all transitions of a signal during an execution. 
The \textit{Leakage Analyzer} then selects the subset of interesting traces that can lead to a vulnerability.
These traces are  snipped at the last clock cycle at which any signal within a selected module instance toggles. The duration of this constrained trace is then computed as the execution time of the module instance. The \textit{Leakage analyzer} analyses the timing behaviors of each module instance in a DUT to identify the module instance with timing vulnerabilities.

{\flushleft \bf Hierarchical leakage analysis.}
Modern processors, however, consist of hundreds of module instances with various inter-module dependencies~\cite{cva6,boom,rocket_chip_generator}. Analyzing all their timing behaviors is time-consuming and computationally expensive. 
Hence, we require a staggered approach for leakage analysis, which prioritizes modules based on their dependencies. A different module-specific execution time indicates that the source of the vulnerability either originates in the current module or a module lower in the hierarchy.
For example, the memory control unit (MCU) controls the data accesses in the cache. 
Upon storing the data at a memory address, the cache sends the \texttt{Ready} signal to the MCU. Thus, if a timing difference is in the cache delaying the \texttt{Ready} signal, the MCU will also observe the timing difference.

The \textit{Leakage Analyzer} implements a hierarchical method taking a bottom-up approach. In this approach, we map the dependencies between modules and place these modules into various hierarchical levels such that the lower-level modules are sub-instances of a higher-level module. 
The hierarchical leakage analysis is then carried out incrementally from the lowest module to the highest module. For example, in \boom{}~\cite{boom} there are 431 module instances, when placed hierarchically they constitute 10 levels. Thus, \boom{} analysis starts from the tenth level and moves upwards until the top module. This approach also ensures the detection of timing vulnerabilities in all lower levels before detection in a higher level module instance.

Once the \textit{Leakage analyzer} identifies a module instance with execution time differences caused by a pair of inputs, it sends the module instance and the corresponding simulation traces to \textit{Diagnozer} to pinpoint the location of timing differences further. 

\subsection{Localizing the Source of Timing Vulnerabilities}\label{sec:diagnozer}
For a pair of inputs exhibiting a difference in execution time in a particular module instance, the \textit{Diagnozer} localizes the source of the timing difference within the design RTL. The \textit{Diagnozer} takes as input two sets of simulation traces, each corresponding to an execution of the DUT with different data inputs. It operates in two phases: (i)~identifying the element causing the divergence and (ii)~mapping the cause in the RTL source code of the processor.

{\flushleft \bf Identifying the element causing the divergence.} 
For the given pair of inputs, the \textit{Leakage Analyzer} generates a set of simulation traces corresponding to each input, $ST_1$ and $ST_2$ for the module under examination, $D$. If each trace in the two sets are exactly identical, then the execution time for the two inputs are equal. On the other hand, if the execution time of the two inputs are different, there exists a subset of the traces within the two sets that differ.
In this subset of simulation traces, the trace which first deviates, corresponds to the combinational element that instigates the timing difference.
Algorithm~\ref{alg:diagnoser} shows the process of the \textit{Diagnozer}. In the first phase, the \textit{Diagnozer} iterates through every clock cycle of simulation traces and finds the first signals which differ in the two sets of traces (Lines 6,7). It creates the list of signals $temp\_{V_s}$ whose traces first differ between $ST_1$ and $ST_2$.

{\flushleft \bf Mapping the cause in the source code.} 
In a module, a timing difference occurs because some hardware elements take varying clock cycles based on the input. The Sequential element influences the number of clock pulses for execution.  Hence, to trace the source of the timing difference within the module, the \textit{Diagnozer} traces dependencies from the identified combinational signals to the sequential nodes in the MEG.\\
In Algorithm~\ref{alg:diagnoser}, this tracing is done by a Breadth First Search to identify the subsequent sequential elements which originate from each signal in $temp\_V_s$. This is stored in the list $V_s$ along with line numbers where the corresponding event on the sequential element is found in the RTL of $D$.

\begin{algorithm}[t]
\SetFuncSty{textsc}
\SetKwInOut{KwIn}{Input}
\SetKwInOut{KwOut}{Output}
\DontPrintSemicolon
\caption{The Diagnozer in \ourtool{} to locate the timing vulnerabilities in the DUT.}\label{alg:diagnoser}
\KwIn{
    $(ST_1,ST_2)$ \tcp{Pair of simulation traces}
    $\mathbb G(D)$ \tcp{MEG of  module under examination}
}
\KwOut{
    $(V_s, L)$ \tcp{Set of signals identified as the cause of the timing difference and corresponding line numbers in RTL,}

}

$V_s \gets \phi$; $temp\_V_s \gets \phi$; $L \gets \phi$\;
\tcc{Phase 1}
\For{ every clock cycle ($clk$) in $ST_1$}{
    \For{$s \in \mathbb G(D) \cdot S$}{
        \If{$s \in ST_1[clk] \neq s \in ST_2[clk]$}{
            $temp\_V_s \gets temp\_V_s \cup \{s\}$\;
        }
    }
    \If{$temp\_V_s$ is not empty}{
    break\;
    }
}

\tcc{Phase 2}

\For{$temp\_V_s$ is not empty}{
    $new\_V_s \gets \phi$\;
    \For{$s \in temp\_V_s$}{
        
        \For{$(s, s_{child}) \in \mathbb G(D) \cdot \Sigma$}{
            \If{$s_{child}$ is sequential}{
                $V_s \gets V_s \cup \{s_{child}\}$\;
                $L \gets L \cup \{line\_no\}$\tcp{RTL line number}
            }
            \Else{
                $new\_V_s \gets new\_V_s \cup \{s_{child}\}$
            }
        }
    
    }
    $temp\_V_s = new\_V_s$
}

return $(V_s, L)$\;

\end{algorithm}

\subsection{Fuzzing Microarchitectural State Space}\label{sec:method_fuzz}

\ourtool{} introduces two components \textit{Operand mutator} and \textit{Coverage analyzer} to generate inputs for timing vulnerability detection and to monitor timing behaviors explored.

\textbf{\noindent{Operand mutator}} consists of specialized mutation algorithms to generate inputs for detecting timing vulnerabilities in processors.
Detecting vulnerabilities requires changing the DUT's timing behaviors by triggering different microarchitectural state transitions~\cite{rajapaksha2023sigfuzz,ibrahim2022microarchitectural,weber2021osiris}. 
Moreover, a valid timing side-channel is data/memory-dependent~\cite{ge2018survey,osvik2006cache,yavuz2022encider,andrysco2015subnormal,wang2014timing}.
Therefore, we constrain \textit{Operand mutator} to mutate only the segments of a test that will change the memory or data values.

Processor fuzzers use sequences of instructions to verify DUTs, as mentioned in Section~\ref{sec:hwfuzzing}. An instruction contains \texttt{opcode} and \texttt{operand} fields~\cite{riscv_home,hennessy2011computer}; both of which cause the microarchitectural state transitions. 
The \texttt{opcode} fields determine the instruction's operation.
Meanwhile, the \textit{operand} fields provide the source and destination registers (i.e., general-purpose registers (GPRs) and control and status registers (CSRs)) and immediate values/memory addresses. 
Some instructions contain immediate memory addresses only (e.g., branch operation instructions). 
To generate data/memory-dependent inputs, we constrain \textit{Operand mutator} to mutate only the immediate values/memory addresses of instructions uniformly at random~\cite{shen2013modern, hennessy2011computer}. To further increase the mutation space, we assign random values to GPRs and valid memory addresses during the processor's initialization. 

However, replacing the original mutation algorithms, \textit{Coverage Mutator}, with \textit{Operand mutator} reduces the fuzzer's efficiency in exploring the design space for timing vulnerabilities. 
This is because the mutator will not change the \texttt{opcode} and register-operands of instructions. 
Therefore, to maintain the efficacy of design space exploration and guarantee the effectiveness of timing vulnerability detection simultaneously, we use \textit{Coverage mutator} to explore design spaces and use \textit{Operand mutator} to exploit time side-channels near the design spaces explored. 
Any tests achieving new code coverage will be identified as seeds for \textit{Operand mutator}. 
For each seed, the \textit{Operand mutator} will generate multiple data-dependent inputs, aiming to trigger the microarchitectural state transitions that will cause timing differences compared to the seed.

\textbf{\noindent {Coverage analyzer}} monitors timing behaviors covered by mapping microarchitectural state transitions to executed paths.
For a given DUT, various paths exist between the input nodes and output nodes of the corresponding MEG. For an input value to the DUT, tracing the execution path followed and mapping it to the MEP poses a challenge. 

To address this challenge, we utilize \systemverilog{} Assertion (SVA) properties known as \texttt{cover} properties~\cite{IEEEstd}. 
If an \texttt{cover} property evaluates to true for a given input, the DUT enters a stage during simulation when the property holds. 
Each graphical path is converted to a \texttt{cover} property using the annotated edge conditions and the timing behavior of the node. 
The timing behavior of a node depends upon the type of node defined. 
Algorithm~\ref{alg:cov_ana} shows the process of generating the conditions of a MEP, $P \in \mathbb G(D)$, as the expression of a \texttt{cover} property.
As combinational nodes are modeled after combinational logic, events occurring on these nodes complete instantaneously (line 4), while those occurring on sequential nodes complete on the next clock edge (line 6). 
If the value of a signal is from other modules, including input signals and signals connected with subinstances, we assume the events will \textit{eventually} happen (line 8). 
Appendix~\ref{apd:sva_case} shows the \texttt{cover} properties for MEPs in the example cache set. 

\begin{algorithm}[t]
\SetFuncSty{textsc}
\SetKwInOut{KwIn}{Input}
\SetKwInOut{KwOut}{Output}
\DontPrintSemicolon
\caption{Coverage analyzer}\label{alg:cov_ana}
\KwIn{
    $P$ \tcp{Micro-Event Path}
    $\mathbb G(D)$ \tcp{MEG of module $D$}
}
\KwOut{
    $Condition$ \tcp{Assertion Property corresponding to Micro-Event Path $P$}
}

$Condition \gets \left \langle  \right \rangle$ \tcp{Empty Sequence}
\For{$(s_{i1}, s_{i2}) \in P$}{
    \If{$(s_{i1}, s_{i2})$ \textbf{has} $branch$}{
        $Condition \gets Condition \blue{||}\red{\cup \{branch\}} \blue{\left \langle branch \right \rangle}$\;
    }
    \If{$s_{i2} \in S_q$}{
        $Condition \gets Condition \blue{||}\red{\cup\{1~cycle\}} \blue{\left \langle 1~cycle \right \rangle}$\;
    }
    \If{$s_{si1} \in S_I$}{
        $Condition \gets Condition \blue{||}\red{\cup\{eventually\}}\blue{\left \langle eventually \right \rangle}$\;
    }
}

return $Condition$\;

\end{algorithm}

Properties representing all possible paths in the MEG corresponding to the module are instrumented in the RTL. The \textit{DUT Simulation} unit of the Vulnerability Detection module takes as input this modified RTL. The \textit{Coverage Analyzer} utilizes the results of the resultant assertion report for calculating the coverage of the various graphical paths.

\section{Evaluation}\label{sec:experiment}
We evaluate \ourtool{} on three most advanced open-sourced processors based on ~\riscv{}~\cite{riscv_home} instruction set architecture~(ISA). 
We first demonstrate the new vulnerabilities detected by \ourtool{} and provide statistical analysis to prove the existence of the timing side channels.
We then leverage the power of \graph{} of \ourtool{} to identify the root causes of these vulnerabilities and evaluate the efficiency of our framework, as shown in Table~\ref{tab:DiagRes}.  Finally, we evaluate the timing behaviors covered by fuzzing.

\subsection{Evaluation Setup}
\textbf{Benchmark selection.} Most commercial processors are protected intellectual properties without available source code. 
Thus, we pick the three large (in terms of the number of gates) and widely-used open-sourced processors: \rc{}~\cite{rocket_chip_generator}, \boom{}~\cite{boom}, and \cva{}~\cite{cva6} from the \riscv{} ISA.
Most recent hardware security tools are evaluated using these processors~\cite{kandethehuzz,hur2021difuzzrtl,chen2023hypfuzz}.
The \cva{} and \boom{} are more complex compared to the \rc{}. 
They possess advanced micro-architectural features such as out-of-order execution and support single instruction-multiple data (SIMD) execution.

\noindent\textbf{Evaluation environment.} We use the industry-standard tool, Synopsys \vcs{}~\cite{vcs} for DUT simulation. We convert timing behaviors of RTL modules into \systemverilog{} Assertion~(SVA) \texttt{cover} properties and instrument them into DUTs. 
We use \textit{Chipyard}~\cite{chipyard} environment for the processors. We collect the coverage report and simulation traces from \vcs{} to analyze the timing coverage and timing behaviors, respectively.

\noindent\textbf{Fuzzing setup.} We use \hypfuzz{}~\cite{chen2023hypfuzz} to generate the seeds for \textit{Operand mutator}. Other processor fuzzers that generate sequences of instructions as inputs are also compatible~\cite{rfuzz,hur2021difuzzrtl,kandethehuzz,xu2023morfuzz,chen2023psofuzz, solt2024cascade, gohil2023mabfuzz}.
\hypfuzz{} is a state-of-the-art hardware fuzzer that combines fuzzing and formal tools to maximize coverage and speed up design exploration.
\hypfuzz{} is also compatible with various coverage metrics. 
We use a combination of branch, condition, and FSM metrics for code coverage.  
Branch and condition metrics monitor the combinational logic of DUTs. The FSM metric monitors the sequential logic of DUTs~\cite{kandethehuzz,chen2023hypfuzz,vcs}. 
Therefore, any new points covered by inputs represent at least one new microarchitectural state transition triggered. 
We collect these inputs as seeds and use the \textit{Operand mutator} to generate 200 data-dependent inputs for each seed. 
We ran the entire fuzzing process for 72 hours, and repeated it thrice to collect coverage results.

\begin{figure*}

    \begin{subfigure}{.32\textwidth}
      \centering
      \includegraphics[width=0.8\linewidth]{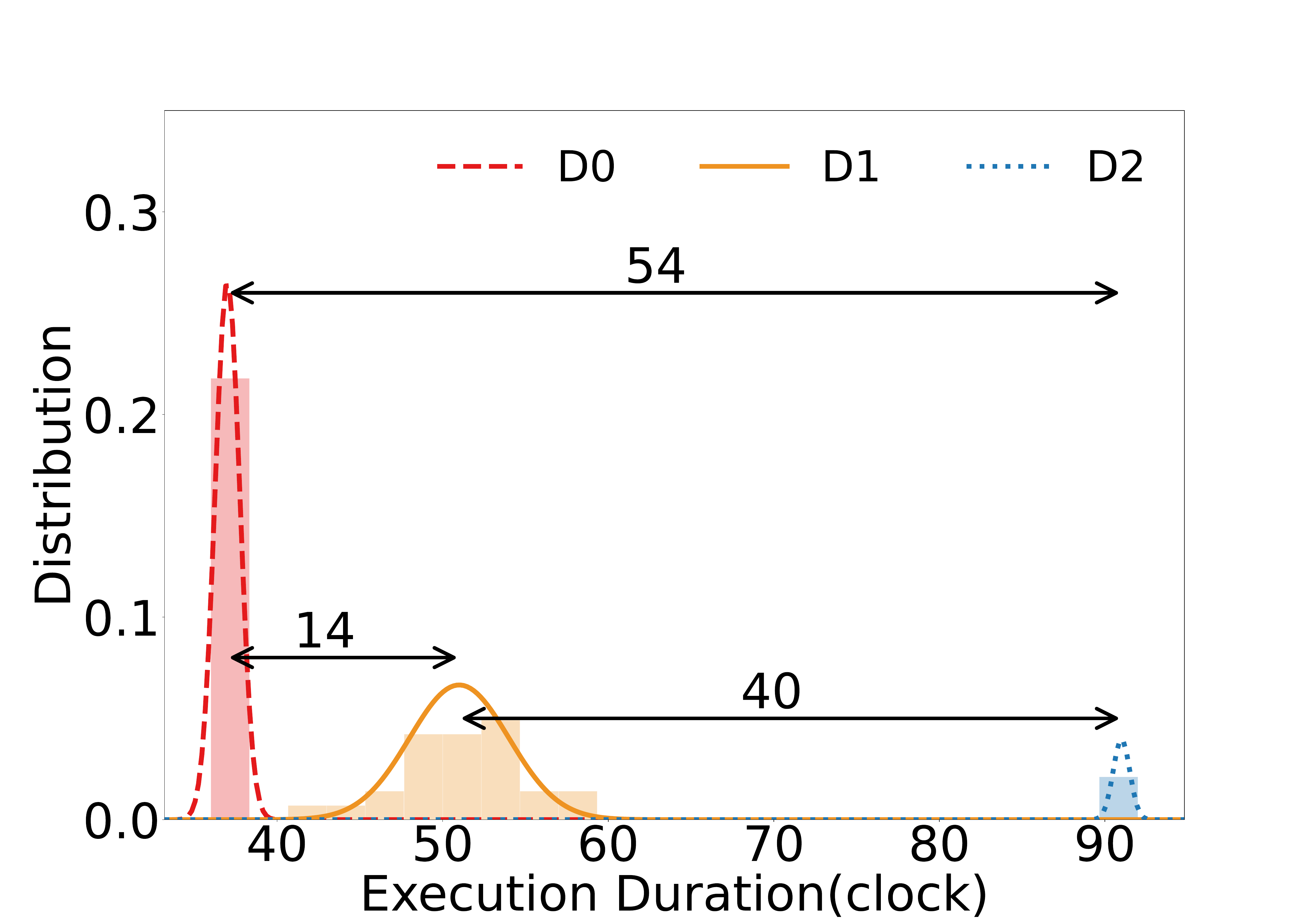}
      \caption{\cva{}~\cite{cva6}: \texttt{DIVUW}}
      \label{fig:cva_divuw}
    \end{subfigure}%
    \hfill{}
    \begin{subfigure}{.32\textwidth}
      \centering
      \includegraphics[width=0.8\linewidth]{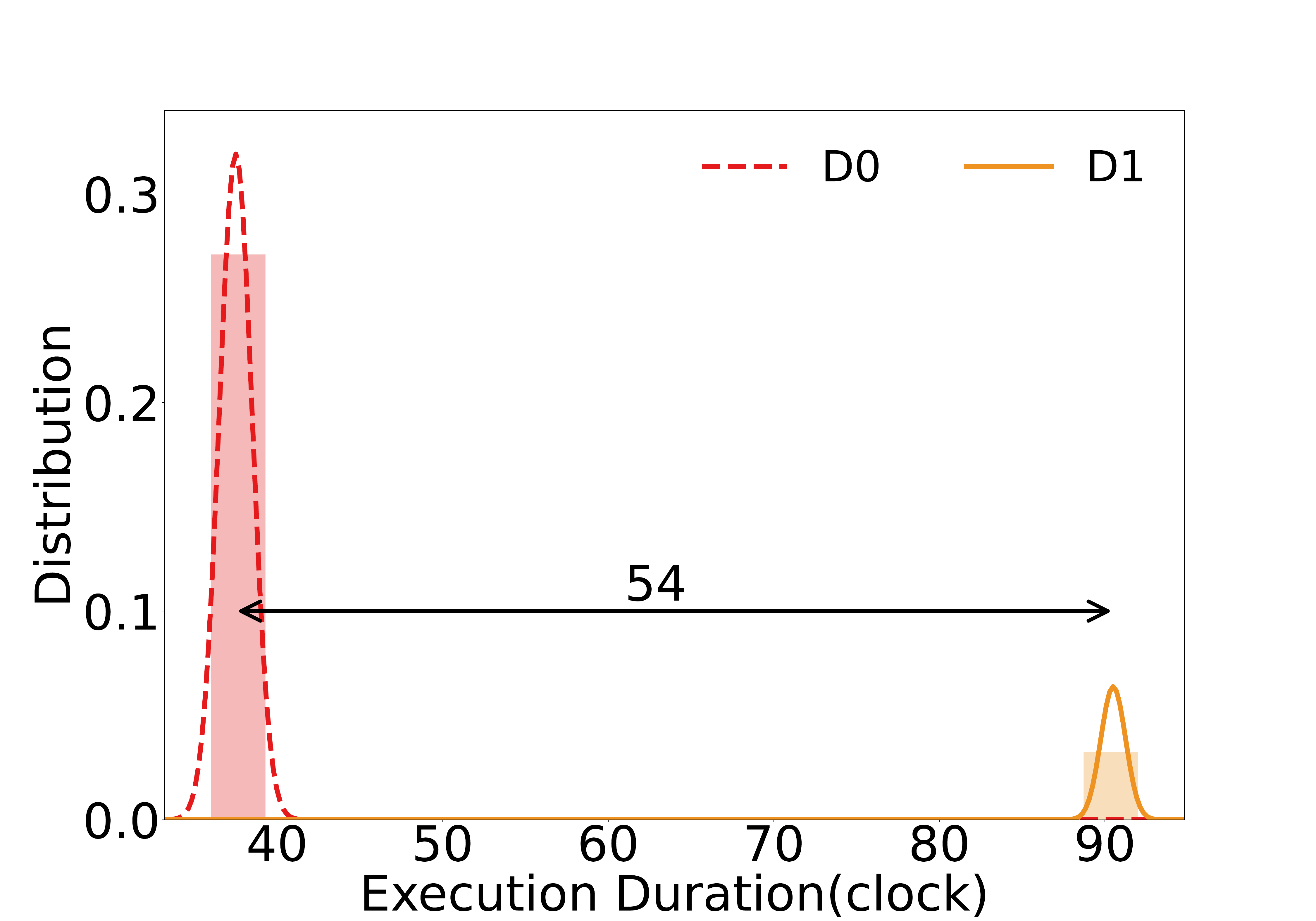}
      \caption{\cva{}~\cite{cva6}: \texttt{REMW}}
      \label{fig:cva_remw}
    \end{subfigure}%
    \hfill{}
    \begin{subfigure}{.32\textwidth}
      \centering
      \includegraphics[width=0.8\linewidth]{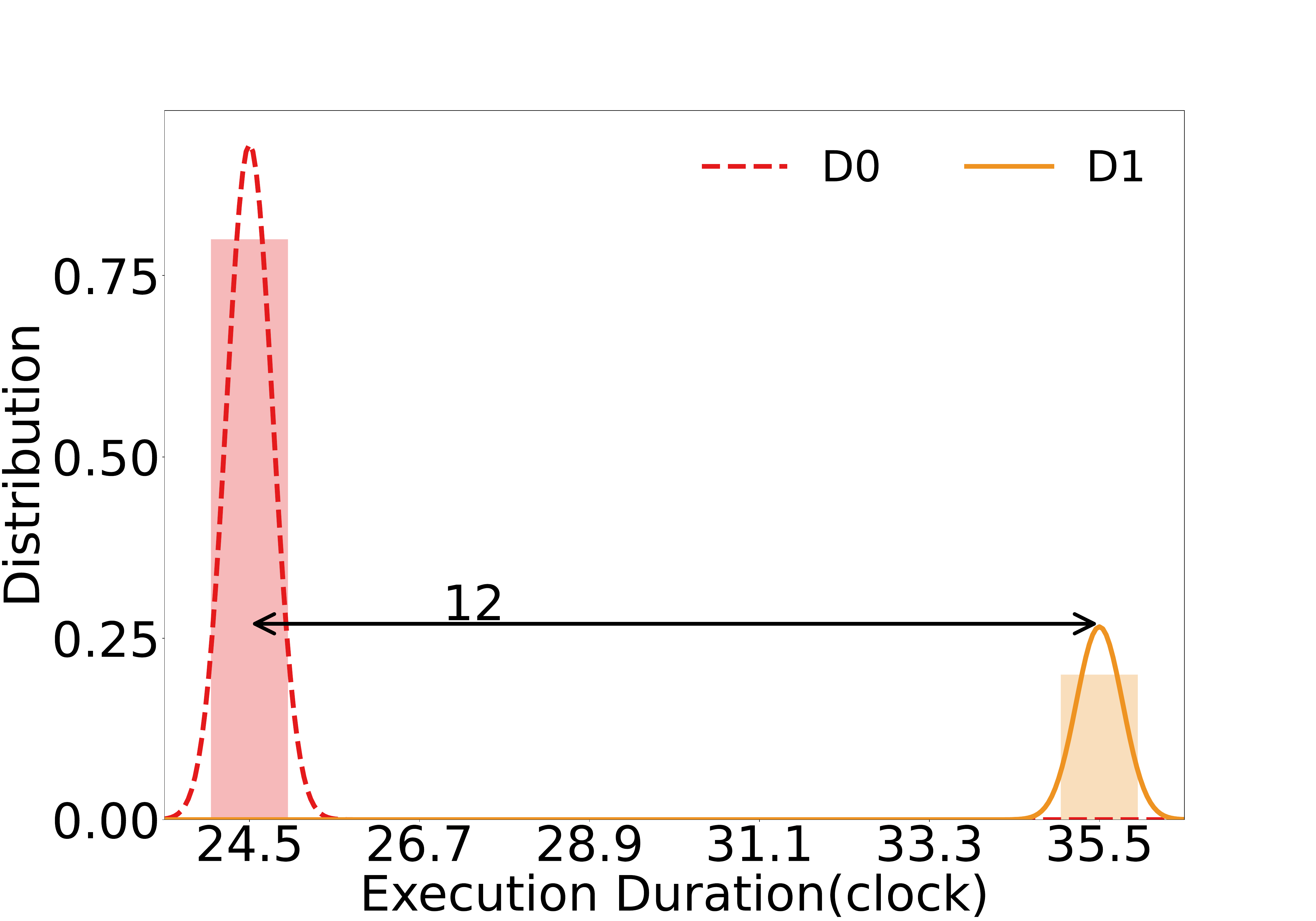}
      \caption{\cva{}~\cite{cva6}: \texttt{C.MV,MV}}
      \label{fig:cva_mv}
    \end{subfigure}%
    \hfill{}
    \begin{subfigure}{.32\textwidth}
      \centering
      \includegraphics[width=0.8\linewidth]{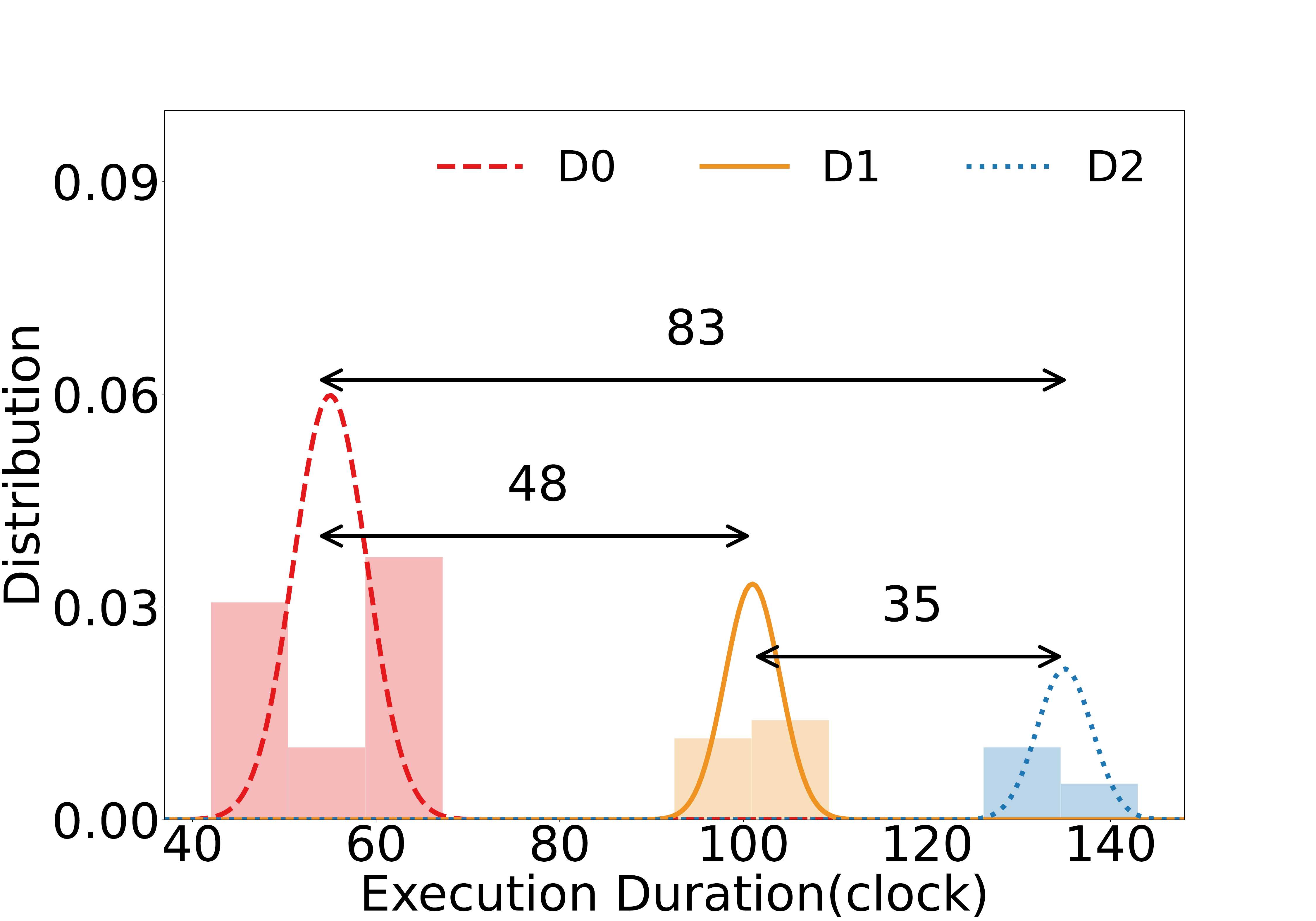}
      \caption{\boom{}\cite{boom}: \texttt{DIVUW + REM}}
      \label{fig:boom_divuw_rem}
    \end{subfigure}%
    \hfill{}
    \begin{subfigure}{.32\textwidth}
      \centering
      \includegraphics[width=0.8\linewidth]{figures/other_cva6.pdf}
      \caption{\cva{}~\cite{cva6}:
      \texttt{C.ADD[W],C.SUB[W]}}
      \label{fig:cva_add_sub}
    \end{subfigure}%
    \hfill{}
    \begin{subfigure}{.32\textwidth}
      \centering
      \includegraphics[width=0.8\linewidth]{figures/other_cva6.pdf}
      \caption{\cva{}~\cite{cva6}: \texttt{C.AND,C.OR,C.XOR}}
      \label{fig:cva_and_or}
    \end{subfigure}

    \centering
    \caption{Timing behaviour of detected novel side-channels.}
    \label{fig:timing_behaviour}
\end{figure*}

\subsection{Detecting Novel Side Channels}\label{sec:exp_novel}
This section will discuss 12 new timing vulnerabilities found by \ourtool{}. Furthermore, Appendix~\ref{apd:poc_trigger_code} contains the proof of concept code for mentioned vulnerabilities.

\noindent\textbf{\texttt{DIVUW + REM} in \boom{}~\cite{boom}.} The side channel under consideration pertains to the consecutive execution of the \texttt{DIVUW} and \texttt{REM} instructions. In Figure \ref{fig:boom_divuw_rem}, the operational characteristics of this side channel are graphically depicted across various operand executions. Our analysis revealed median discrepancies of 48 cycles, 35 cycles, and 83 cycles between the divisor equal to 0, 1, and greater than 1 distributions respectively, with a maximum deviation of 101 cycles observed when deliberately selecting operands to maximize the disparity. Establishing a threshold for the median separation facilitates a successful discrimination rate of 100\% when distinguishing between binary states 0 and 1.

{\noindent\textbf{\texttt{DIVUW} in \cva{}~\cite{cva6}.}} The side channel pertains to the execution of \texttt{DIVUW} instruction. Figure \ref{fig:cva_divuw} illustrates the operational characteristics of this side channel across multiple operand executions. Notably, our analysis has unveiled median discrepancies of 14 cycles, 39 cycles, and 54 cycles among the divisor equal to 0, 1, and greater than 1 distributions respectively. Furthermore, we have observed a maximum deviation of 56 cycles when deliberately selecting divide by zero to maximize the disparity.
Establishing a threshold for the median separation enables the successful discrimination of binary states 0 and 1 with a 100\% accuracy rate. However, our findings indicate that attackers leverage this channel to transmit three states rather than the intended two states by defining two thresholds, achieving a success rate of 91\%. \\
{\noindent\textbf{\texttt{REMW} in \cva{}~\cite{cva6}.} }\ourtool{} detected this novel side channel when fuzzing \cva{} with \texttt{REMW} instruction.
Figure \ref{fig:cva_remw} shows the operational characteristics of this side channel across multiple instances of operand executions. We demonstrate a median difference of 54 cycles between the divisor equal to 0 and greater than 0 timing behavior distributions. Additionally, when we select zero as a divisor in the \texttt{REMW} instruction to maximize the timing difference, the maximum deviation is 56 cycles. Attackers can use this channel by 
establishing a threshold for the median separation to transmit binary states 0 and 1 with 100\% accuracy.

{\noindent\textbf{\texttt{C.ADD[W],~C.SUB[W],~C.AND,~C.OR,~C.XOR,} and~\texttt{[C].MV} in \cva{}~\cite{cva6}.} }\ourtool{}  detected multiple novel side channels that occur when executing compressed \riscv{} instructions, i.e. \texttt{C.ADD[W], C.SUB[W], C.AND, C.OR, C.XOR,} and \texttt{[C.]MV}. \riscv{} Zkt contract has classified these instructions as serious security vulnerabilities if the instructions are data dependent~\cite{zkt_instructions}. The \texttt{MV} instruction has the same timing behavior as its compressed version, i.e., \texttt{C.MV} and causes a timing channel.
Figures \ref{fig:cva_add_sub}, \ref{fig:cva_and_or}, and \ref{fig:cva_mv}, show the operational characteristics of these side channels across multiple instances of operand executions. We demonstrate a median difference of 12 cycles between the second operand equal to 0 and greater than 0 timing behavior distributions. 
Selecting a specific value of
zero as the operands of these instructions results in 12 more cycles compared to any other operand values.
Establishing a threshold for the median separation allows for the successful discrimination between binary states 0 and 1 with an accuracy rate of 100\%.

\subsection{Redetecting Known Side Channels}\label{sec:exp_known}
\textbf{\texttt{DIV} in \boom{}~\cite{boom}.} 
\ourtool{} successfully generated test cases that exposed timing side-channel vulnerabilities related to division instructions (\texttt{DIV}), a vulnerability disclosed in \sigfuzz{}~\cite{rajapaksha2023sigfuzz} for the \boom{} processor.
\ourtool{} generated multiple test cases featuring the \texttt{DIV} instruction, and by systematically mutating the input values during the fuzzing process, it revealed variations in the number of clock cycles required for the \texttt{DIV} instruction to complete its operation. Further investigation elucidated that when the divisor was bigger than the dividend, the division unit necessitated more time to conclude the division process.

\noindent\textbf{\texttt{SC} in \rc{}~\cite{rocket_chip_generator}~and~\boom{}~\cite{boom}.} 
\ourtool{} also identified a timing side-channel associated with Store-Conditional operations, effectively diagnosing timing disparities resulting from the presence of the \texttt{dirty} bit in the data cache implementation. 
The vulnerability was discovered in \sigfuzz{}~\cite{rajapaksha2023sigfuzz} for the \rc{} and \boom{} processors. 
The test cases feature a \texttt{SC} instruction containing at least one subsequent load instruction. 
Through the mutation of these test cases, we were able to detect timing discrepancies when the load/store module attempted to access an address not present in the cache. Subsequently, the \textit{Diagnozer} of \ourtool{} pinpointed the root cause of this timing difference, attributing it to the \texttt{dirty} bit that was set by the Store-Conditional instruction for a cache line.

\subsection{Pinpointing the Locations of Side Channels}
We apply \textit{Diagnozer} (See Section~\ref{sec:diagnozer}) to identify the root causes of detected timing vulnerabilities (See Sections~\ref{sec:exp_known},~\ref{sec:exp_novel}). We describe, in detail, the \textit{Diagnozer} results for two novel and two known vulnerabilities. The results for all detected vulnerabilities are summarized in Table~\ref{tab:DiagRes}.

        \bgroup
        \createlinenumber{1}{13512}
        \createlinenumber{2}{13513}
        \createlinenumber{3}{13514}
        \createlinenumber{4}{13515}
        \createlinenumber{5}{13516}
        \createlinenumber{6}{13517}
        \createlinenumber{7}{13518}
        \createlinenumber{8}{13519}
        \createlinenumber{9}{13520}
        \createlinenumber{10}{13521}
        \createlinenumber{11}{13522}
\begin{lstlisting}[language=Verilog, label = {listing:div_cva}, caption={Source location of \textbf{\texttt{DIVUW}} in \cva{}~\cite{cva6}.},style=prettyverilog,float,belowskip=0pt,aboveskip=0pt,firstnumber=1, escapechar=\%,linewidth=\linewidth, xleftmargin=25pt]
    %\Hilight[0.33]%state_q        <= state_d;
    %\Hilight[0.30]%op_a_q         <= op_a_d;
    %\Hilight[0.30]%op_b_q         <= op_b_d;
    res_q          <= res_d;
    %\Hilight[0.27]%cnt_q          <= cnt_d;
    %\Hilight[0.22]%id_q           <= id_d;
    rem_sel_q      <= rem_sel_d;
    %\Hilight[0.44]%comp_inv_q     <= comp_inv_d;
    res_inv_q      <= res_inv_d;
    %\Hilight[0.47]%op_b_zero_q    <= op_b_zero_d;
    %\Hilight[0.57]%div_res_zero_q <= div_res_zero_d;
\end{lstlisting}
        \egroup

{\noindent\textbf{\texttt{DIVUW} in \cva{}~\cite{cva6}.}}
In \cva{}, given the instruction sequence in Listing~\ref{listing:cva6_divuw_poc}, the \textit{Diagnozer} identifies the source of the vulnerability in module  \texttt{serdiv} in the lines highlighted in Listing~\ref{listing:div_cva}. The first phase of the \textit{Diagnozer} identifies 27 signals as the instigating signals, while the second phase pinpoints that these signals change the values of 8 sequential signals.

{\noindent\textbf{Compressed Instructions} in \cva{}~\cite{cva6}.}
Given two trace files corresponding to the instruction sequence in Listing~\ref{listing:cva6_other_poc} the first phase of the \textit{Diagnozer} identifies the sources of these vulnerabilities in the \texttt{ALU} module. The exact RTL lines are as highlighted in Listing~\ref{listing:ci_cva}. Though the \texttt{ALU} module does not have a sequential component, the effects of its outputs are propagated to the inputs of the other modules thereby influencing further execution. \\

{\noindent\textbf{\texttt{DIVUW + REM}} in \boom{}~\cite{boom} and \rc{}~\cite{rocket_chip_generator}.}
Given two trace files corresponding to the instruction sequence in Listing~\ref{listing:boom_divuw_rem_poc}, the \textit{Diagnozer} identifies the source of this vulnerability in \boom{}~\cite{boom} and \rc{}~\cite{rocket_chip_generator} at different lines in module  \texttt{MulDiv}. The vulnerability is localized to the sequential elements \texttt{divisor, state, negout} in both processors (refer Table~\ref{tab:DiagRes}).

{\noindent\textbf{Division by zero} in \boom{}~\cite{boom}, \rc{}~\cite{rocket_chip_generator}, \cva{}~\cite{cva6}.}
Given two trace files corresponding to the instruction sequence in Listing~\ref{listing:boom_divuw_rem_poc} with the divisor set to 0, the \textit{Diagnozer} identifies the sources of this vulnerability in \boom{}, \rc{}, \cva{} as shown in Table~\ref{tab:DiagRes}. 
Though the root cause is localized to the same module in \rc{} and \boom{}, the pinpointed lines differ. While in \cva{}, the divide by zero vulnerability is localized to module \texttt{serdiv}.

Hence, though the same timing vulnerability can affect multiple DUTs, due to the differences within the microarchitectural design, the source of the vulnerability in the RTL code differs. Furthermore, consider the two division-related vulnerabilities detected in \cva{}~\cite{cva6}. Although these vulnerabilities affect the same module (\texttt{serdiv}), the source of the vulnerabilities within the module differs. Hence, the automated localization of vulnerability sources performed by \ourtool{} is beneficial and efficient.

\begin{figure*}[h]
    \begin{subfigure}{.32\textwidth}
      \centering
      \includegraphics[width=1\linewidth]{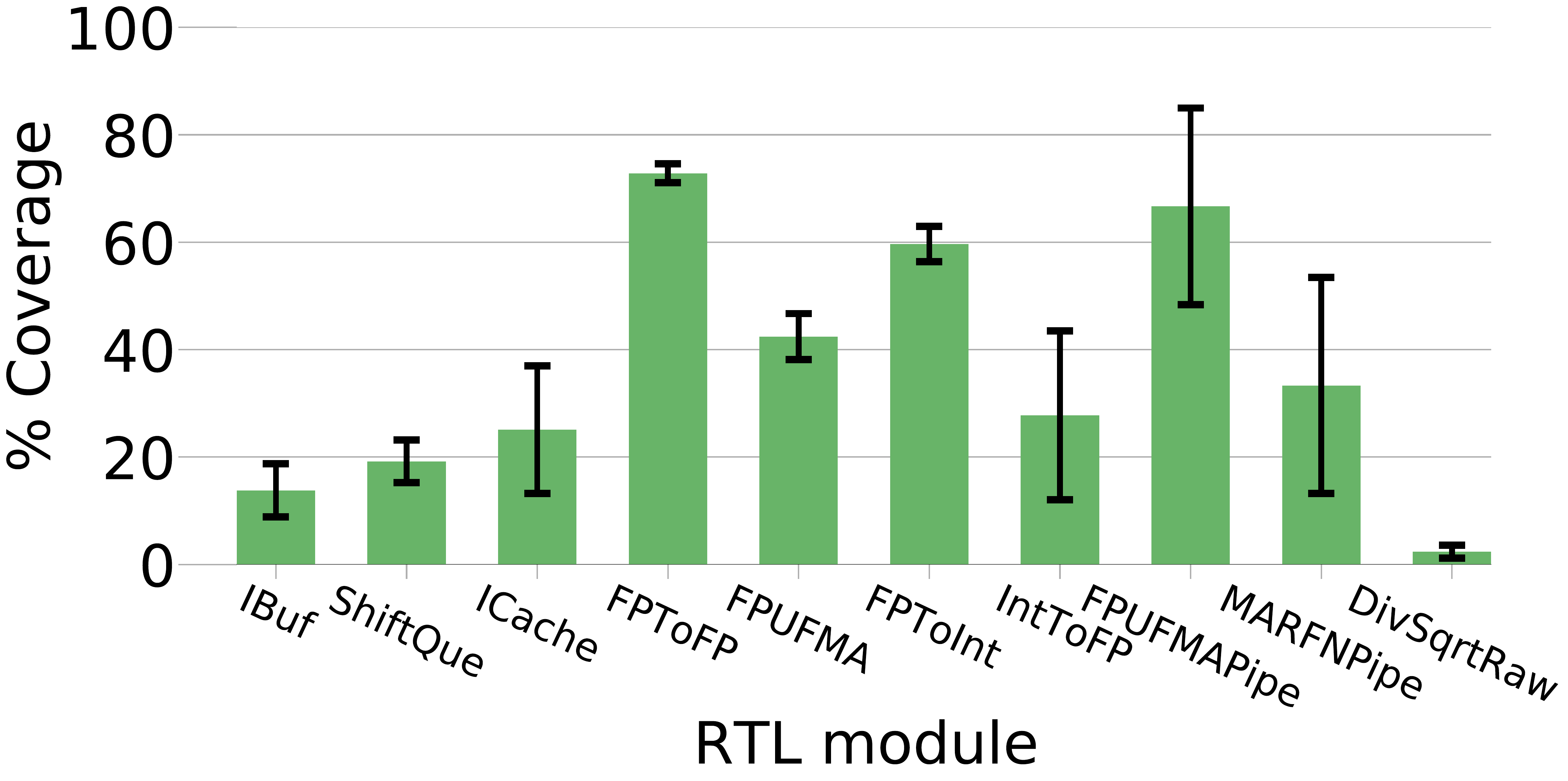}
      \caption{ \rc{}~\cite{rocket_chip_generator}}
      \label{fig:rocket_cov}
    \end{subfigure}%
    \hfill{}
    \begin{subfigure}{.32\textwidth}
      \centering
      \includegraphics[width=1\linewidth]{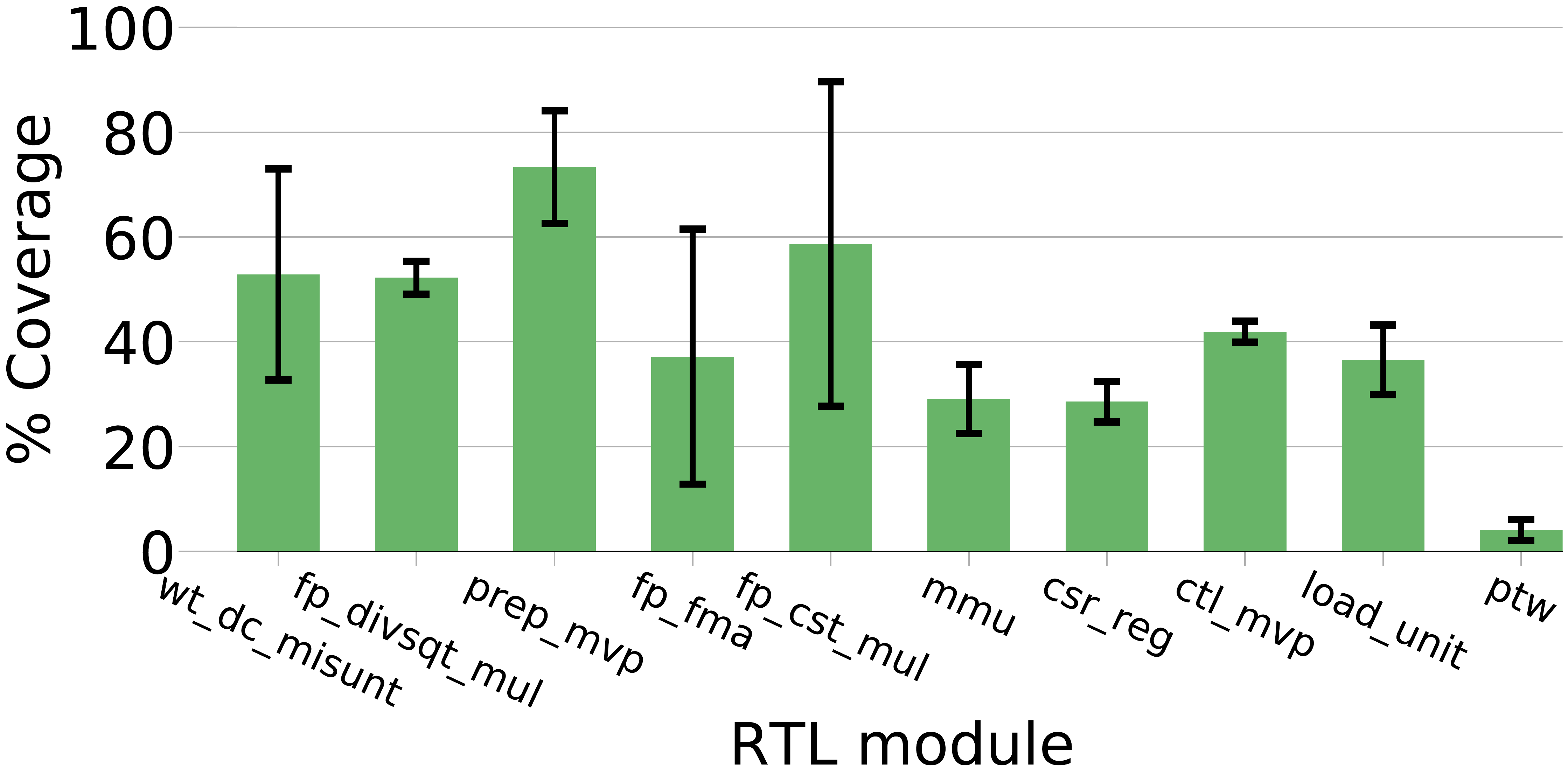}
      \caption{\cva{}~\cite{cva6}}
      \label{fig:cva_cov}
    \end{subfigure}
    \hfill{}
    \begin{subfigure}{.32\textwidth}
      \centering
      \includegraphics[width=1\linewidth]{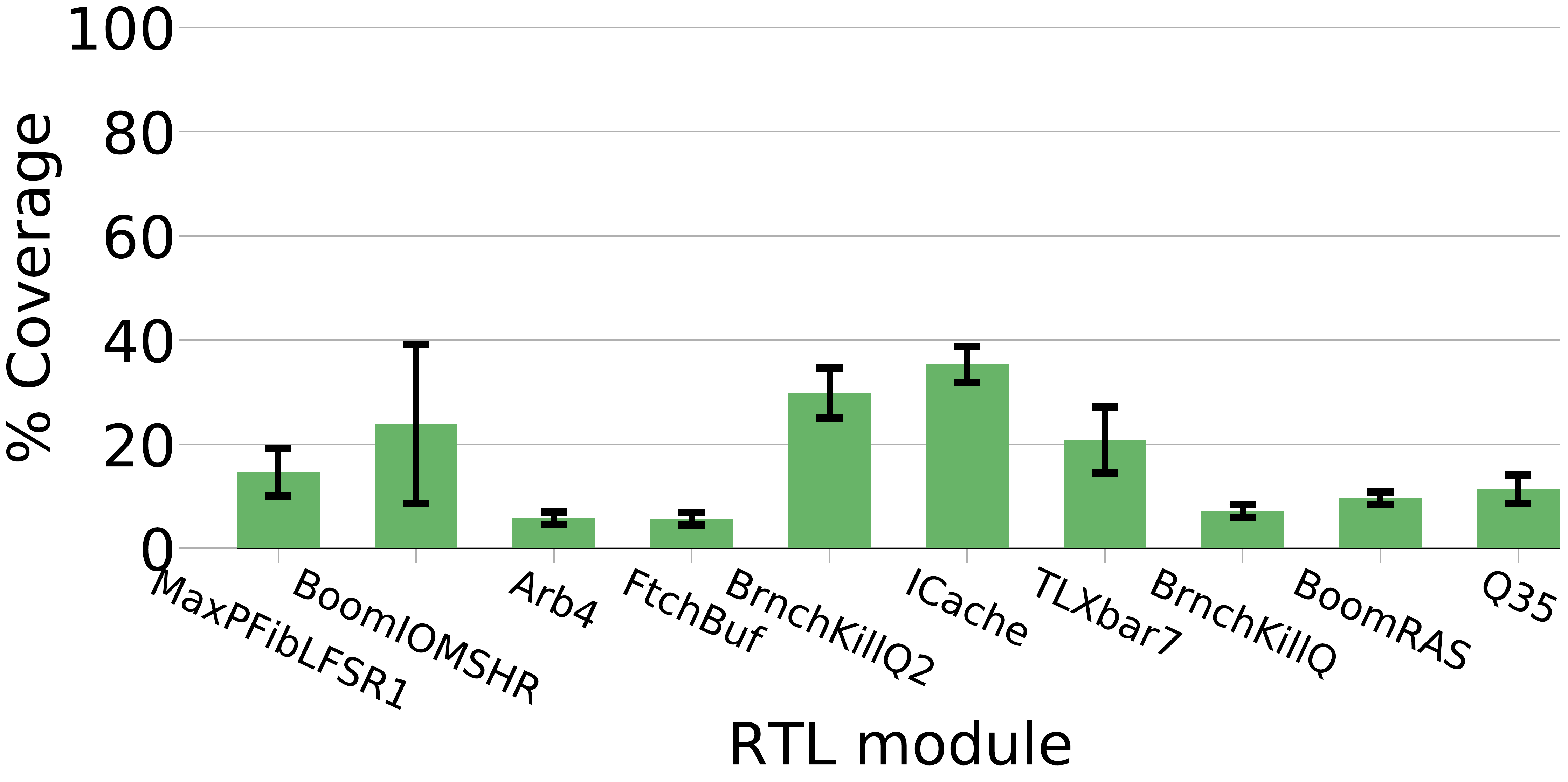}
      \caption{\boom{}~\cite{boom}}
      \label{fig:boom_cov}
    \end{subfigure}
    \centering
    \caption{Timing coverage of RTL modules in \cva{}~\cite{cva6}, \rc{}~\cite{rocket_chip_generator}, and \boom{}~\cite{boom}. The black line indicates the variation in coverage offered.\vspace{-0.5cm}}
    \label{fig:timing_cov}
\end{figure*}

\subsection{Coverage Analysis}

We use \textit{Coverage analyzer} to monitor timing behaviors explored by inputs from \textit{Operand mutator}. 
\textit{Coverage analyzer} converts paths of \graph{} of an RTL module into \texttt{cover} properties and instrument these properties in to DUT. 
After simulation, \textit{Coverage analyzer} collects assertion results to calculate the timing behaviors covered (See Section~\ref{sec:method_fuzz}).
Figure~\ref{fig:timing_cov} shows the coverage achieved in various modules of the three benchmarks. 
Due to the space limitation, we only show the coverage of 10 modules with the highest total number of timing coverage points. 
On average, \ourtool{} achieves $39.57\%$, $33.16\%$, and $20.20\%$ coverage on \cva{}~\cite{cva6}, \rc{}~\cite{rocket_chip_generator}, and \boom{}~\cite{boom}, respectively. 

Since we emphasize covering timing behaviors, our \textit{Operand mutator} generates 200 inputs for each seed generated by the coverage-feedback fuzzer. 
The overall coverage of \ourtool{} can be increased by using more seeds from coverage-feedback fuzzer and running the fuzzer for more time. 
Also, many coverage points are left uncovered due to the multiple configurations of DUT.
For example, \cva{} has parameters to configure its float-point unit to support operations from 8-bit to 128-bit~\cite{cva6}. 
\ourtool{} fuzzes \cva{} in its default configuration where
\cva{} uses 64-bit operations. 
Hence, all other operations' timing behaviors are uncoverable. \rc{}'s floating-point unit~\cite{rocket_chip_generator} and \boom{}'s branch predictor~\cite{boom} have similar configurations. Therefore, these configurations reduce the overall coverage. 

\subsection{Exploitability of Detected Vulnerabilities}
In this section, we discuss the potential exploitations of the vulnerabilities detected by \ourtool{} as presented in Section~\ref{sec:experiment}. Such vulnerabilities can be exploited for information leakage across diverse scenarios as described below. 

{\noindent  \textbf{Covert Channels.}} A timing covert channel breaks the process isolation guarantees provided by the hardware. A \textit{sender} process can perform operations influencing the execution time of a \textit{receiver} process, which infers a bit value based on this observed timing. For instance, the \texttt{DIVUW}-based vulnerability detected by \ourtool{} in \cva{}~\cite{cva6} can be employed to design a covert channel based on the timing differences. However, realizing such a covert channel requires the communicating processes to execute on the physical core using hyperthreading features which are unavailable on our evaluation processors\cite{rocket_chip_generator,boom,cva6}.

{\noindent \bf Speculative Execution Attacks.} Such attacks happen in out-of-order processors when, during the rolled back of speculatively executed instructions, processor leave their footprints on the micro-architectural components such as the cache. This has been exploited in several popular attacks ~\cite{boom_attacks,Kocher2018spectre,Lipp2018meltdown,rajapaksha2023sigfuzz}. An attacker can formulate a similar attack with the vulnerabilities found by \ourtool{}. For instance, with speculative execution support on \cva{}, a combination of the load and time-dependent instructions(any instruction that is detected by \ourtool{}, See Section \ref{sec:exp_novel}) can utilized to encode sensitive data into the cache, which a cache timing attack can then glean. However, the current state of the evaluated processors is limited to non-speculative execution.

{\noindent \bf Attacking Library Implementations.} An adversary can exploit the timing differences based on the operand values discovered by \ourtool{}  to glean sensitive information from popular libraries in cryptography or machine learning domains. For such an attack, the library implementation is required to have the same instruction flow, e.g., \texttt{DIVUW} followed by \texttt{REM} for \boom{}~\cite{boom} or \texttt{C.ADD} for \cva{}\cite{cva6} with the operand dependent on a secret value.

\section{Related Work}\label{sec:related_work} 

\begin{table}[]
\centering
\caption{Comparison with prior works on timing vulnerability detection on processors with \ourtool{}.  (\textit{N.A.: Not applicable, TSC: Timing side channel.})}
\resizebox{\linewidth}{!}{%
\begin{tabular}{|M{0.27\columnwidth}|M{0.08\columnwidth}|M{0.06\columnwidth}|M{0.18\columnwidth}|M{0.3\columnwidth}|M{0.06\columnwidth}|M{0.08\columnwidth}|M{0.08\columnwidth}|}
\hline
\rotatebox{90}{\textbf{Paper}}    & \rotatebox{90}{\textbf{\begin{tabular}[c]{@{}c@{}}Manual\\ effort\end{tabular}}} & \rotatebox{90}{\textbf{Scalable}} & \rotatebox{90}{\textbf{\begin{tabular}[c]{@{}c@{}}Design\\ source\end{tabular}}}                                             & \rotatebox{90}{\textbf{\begin{tabular}[c]{@{}c@{}} Timing\\ vulnerability\end{tabular}}}        & \rotatebox{90}{\textbf{Coverage}} & \rotatebox{90}{\textbf{\begin{tabular}[c]{@{}c@{}}Root cause\\ analysis\end{tabular}}}\\ \hline
\textbf{\upec{}~\cite{fadiheh2019processor}}           & \tikzcmark{}                                                         & \tikzxmark{}                    & RTL                                                                & Covert Channels                                                                         & N.A.              & \tikzxmark{} \\ \hline
\textbf{\fadiheh{}~\cite{fadiheh2020formal}} & \tikzcmark{}                                                         & \tikzxmark{}                    & RTL                                                                & Covert Channels                                                                         & N.A.              & \tikzxmark{} \\ \hline
\textbf{\checkmate{}~\cite{trippel2018checkmate}}      & \tikzcmark{}                                                         & N.A.                    & \begin{tabular}[c]{@{}c@{}}Abstract\\ model\end{tabular} & Cache TSC                    & N.A.             & \tikzxmark{} \\ \hline
\textbf{\osiris{}~\cite{weber2021osiris}}         & \tikzxmark{}                                                     & \tikzcmark{}                    & Black-box                                                          & \begin{tabular}[c]{@{}c@{}}Eviction-based\\ TSC\end{tabular} & N.A.                 & \tikzxmark{} \\ \hline
\textbf{\absynthe{}~\cite{gras2020absynthe}}       & \tikzxmark{}                                                     & \tikzcmark{}                    & Black-box                                                          & \begin{tabular}[c]{@{}c@{}}Contention-based\\ TSC\end{tabular}               & N.A.                 & \tikzxmark{} \\ \hline
\textbf{\plumber{}~\cite{ibrahim2022microarchitectural}}        & \tikzcmark{}                                                         & \tikzcmark{}                    & Black-box                                                          & \begin{tabular}[c]{@{}c@{}}Variants of cache \\ TSC\end{tabular}       & N.A.                 & \tikzxmark{} \\ \hline
\textbf{\sigfuzz{}~\cite{rajapaksha2023sigfuzz}}        & \tikzxmark{}                                                     & \tikzcmark{}                    & RTL                                                                & TSC                                                                    & \tikzxmark{}                 & \tikzxmark{} \\ \hline
\textbf{\ourtool{}}        & \tikzxmark{}                                                     & \tikzcmark{}                    & RTL                                                                & TSC                                                                    & \tikzcmark{}                 & \tikzcmark{} \\ \hline

\end{tabular}
}
\label{tab:StraDiff}
\end{table}

Existing state-of-the-art techniques for timing side-channel vulnerability detection primarily employ  
formal approaches ~\cite{fadiheh2020formal,fadiheh2019processor,trippel2018checkmate, gleissenthall2019iodine, wang2023specification} and fuzzing ~\cite{weber2021osiris,ibrahim2022microarchitectural,gras2020absynthe,rajapaksha2023sigfuzz} techniques. However, these approaches still exhibit critical shortcomings. In contrast to \ourtool{}, these approaches fail to pinpoint the root causes of the detected timing vulnerabilities without manual efforts that take a long time. Thus, the mitigation based on these techniques is coarse-grained rendering them inefficient in terms of the computational resources in DUT. Further, the coverage metrics used by these solutions, such as hardware performance counters or code coverage~\cite{rajapaksha2023sigfuzz, weber2021osiris, ibrahim2022microarchitectural} do not capture the timing behaviors of the DUT, resulting in uncertainty prior to tape-out~\cite{verifiwhitepaper,gopinath2014code,ivankovic2019code}. In this section, we discuss these perform a comparative analysis with \ourtool{}, as illustrated in Table~\ref{tab:StraDiff}.

\noindent\textbf{Formal approaches for timing vulnerability detection.} 
\textbf{\upec{}}~\cite{fadiheh2019processor,fadiheh2020formal} is a white-box approach to detect side channels in RISC-V RTL designs using SAT-based bounded model-checking. However, such an approach is not scalable to complex processor designs. Alternatively, \textbf{\checkmate{}}~\cite{trippel2018checkmate} 
employs \textit{micro-happens-before} graphs to analyze transient execution vulnerabilities and timing side channels. It detects patterns within these graphs to assess the susceptibility of architectural models to timing side-channel threats. In contrast to our methodology, \checkmate{} relies on matching patterns of vulnerable instructions, while 
\ourtool{} is semantically oriented and automatic.

\noindent\textbf{Fuzzing-based approaches for timing vulnerability detection.}
\textbf{\osiris{}}~\cite{weber2021osiris} is a black-box fuzzer that identifies timing vulnerabilities in commercial processors by brute-forcing different combinations of instruction sequences. However, to reduce the search space, it limits the instruction sequence length to one, leaving vulnerabilities requiring multiple instructions~\cite{percival2005cache} or specific operands~\cite{osvik2006cache} to trigger undetected. 
\textbf{\absynthe{}}~\cite{gras2020absynthe} 
and\textbf{~\plumber{}}~\cite{ibrahim2022microarchitectural}
identify combinations of instructions that trigger microarchitectural timing side-channel leakages by deriving a \textit{leakage template}. However, \plumber{} requires manual efforts to specify mutation algorithms and potential behaviors of a DUT to generate this template. Further, it is limited to the existing cache module and cannot locate them in the DUT.

{\noindent \textbf{\sigfuzz{}}}~\cite{rajapaksha2023sigfuzz} is a grey-box fuzzer that detects the existence of timing vulnerabilities in processors at the RTL.
It generates combinations of instructions to identify cycle-accurate microarchitectural timing side-channels. 
However, replacing instructions can create additional architectural differences, such as differences in general-purpose registers, resulting in a high rate of false positives. Further, \textbf{\textit{SIGFuzz}} suffers from limitations in pinpointing vulnerability locations and coverage metrics as the black-box approaches. 

\ourtool{} addresses these limitations of existing works by providing a novel white-box fuzzer with static analysis to detect and pinpoint timing vulnerabilities in executed testcases in processors enabling fine-grained mitigations. \ourtool{} is scalable to complex designs and end-to-end automated with a specialized coverage metric for timing behaviors. 

\section{Discussion}\label{sec:discus}

\textbf{Use of \textit{Code coverage mutator} and \textit{Operand mutator}.} 
\ourtool{} employs both the \textit{Code coverage mutator} and the \textit{Operand mutator} to explore the design space and generate data-dependent inputs, respectively. In our experiments, we set up the use of these two mutators heuristically. The determination of their utilization is an optimization problem that can aid the efficacy of vulnerability detection. We can model the probability of \textit{Operand mutator} covering a timing behavior~\cite{robert2009monte,zhao2019send}. When this probability falls below a threshold, the \textit{Code coverage mutator} can be called to generate new seeds. However, such an analysis is beyond the scope of this paper.

\noindent\textbf{Port scanning vulnerabilities.} 
Contention for a port in the architecture can cause execution delays, enabling attackers to create a high-resolution time side-channel by scanning for port contention~\cite{aldaya2019port, bhattacharyya2019smotherspectre}. However, these attacks depend on the high bandwidth of shared resources and, primarily, the simultaneous multithreading architecture~\cite{tullsen1995simultaneous,aldaya2019port, bhattacharyya2019smotherspectre}. 
The current open-sourced benchmarks lack such advanced architectures~\cite{cva6,rocket_chip_generator,boom}, and hence detecting port scanning is outside the scope of \ourtool{}.

\section{Conclusion}\label{sec:conclu}

Recent hardware fuzzers have showcased their potential to identify timing vulnerabilities in intricate designs, such as processors. 
However, the existing black-box or grey-box fuzzing approaches fall short in pinpointing the precise location or root cause of timing vulnerabilities.  Further, these approaches lack the necessary coverage feedback mechanisms for the exploration of timing behaviors. 
Addressing these gaps, we develop \ourtool{}, the first approach that combines white-box fuzzing with static analysis. 
Its primary objectives are not only to accurately determine the locations of timing vulnerabilities but also to evaluate the timing behaviors. 
\ourtool{} has successfully detected 12 new timing vulnerabilities and all previously known ones in open-source processors. Moreover, it pinpoints the root causes of these vulnerabilities. This opens up novel avenues in vulnerability detection and timely mitigation in processors.

\section{Acknowledgement}
Our research work was partially funded by Intel's Scalable Assurance Program, Deutsche Forschungsgemeinschaft (DFG) – SFB 1119 – 236615297, the European Union under Horizon Europe Programme – Grant Agreement 101070537 – CrossCon, the European Research Council under the ERC Programme - Grant 101055025 - HYDRANOS, the US Office of Naval Research (ONR Award \#N00014-18-1-2058), the Lockheed Martin Corporation, and the Centre for Hardware Security Entrepreneurship Research and Development (C-HERD) project, Ministry of Electronics and Information Technology (MEiTY), Government of India.
This work does not in any way constitute an Intel endorsement of a product or supplier. Any opinions, findings, conclusions, or recommendations expressed herein are those of the authors and do not necessarily reflect those of Intel, the European Union, the European Research Council, Lockheed Martin Corporation, the US Government, or the Indian Government.
\bibliographystyle{plain_auth_limited}
\bibliography{Bibfile.bib}

\begin{thebibliography}{10}

\bibitem{boom_attacks}
{BOOM Speculative Attacks}.
\newblock \url{https://github.com/riscv-boom/boom-attacks}, 2019.
\newblock {Last accessed on 10/01/2023}.

\bibitem{zkt_instructions}
{Zkt "Constant Time" Instruction List}.
\newblock \url{https://github.com/rvkrypto/riscv-zkt-list/blob/main/zkt-list.adoc}, 2021.
\newblock {Last accessed on 10/01/2023}.

\bibitem{vcs}
{Synopsys VCS}.
\newblock \url{https://www.synopsys.com/verification/simulation/vcs.html}, 2022.
\newblock {Last accessed on 10/01/2023}.

\bibitem{zenbleed}
{Cross-Process Information Leak}.
\newblock \url{https://www.amd.com/en/resources/product-security/bulletin/amd-sb-7008.html}, 2023.
\newblock Last accessed on 09/28/2023.

\bibitem{nvd}
{National Vulnerability Database}.
\newblock \url{https://nvd.nist.gov/vuln/search}, 2023.
\newblock Last accessed on 09/28/2023.

\bibitem{aldaya2019port}
A.~C. Aldaya, B.~B. Brumley, et~al.
\newblock {Port Contention for Fun and Profit}.
\newblock {\em IEEE Symposium on Security and Privacy}, 2019.

\bibitem{chipyard}
A.~Amid, D.~Biancolin, et~al.
\newblock {Chipyard: Integrated Design, Simulation, and Implementation Framework for Custom SoCs}.
\newblock {\em IEEE Micro}, 40(4):10--21, 2020.

\bibitem{andrysco2015subnormal}
M.~Andrysco, D.~Kohlbrenner, et~al.
\newblock {On Subnormal Floating Point and Abnormal Timing}.
\newblock {\em IEEE Symposium on Security and Privacy}, 2015.

\bibitem{rocket_chip_generator}
K.~Asanović, R.~Avizienis, et~al.
\newblock {The Rocket Chip Generator}.
\newblock (UCB/EECS-2016-17), Apr 2016.

\bibitem{baier2008principles}
C.~Baier and J.-P. Katoen.
\newblock {Principles of Model Checking}.
\newblock 2008.

\bibitem{bernstein2005cache}
D.~J. Bernstein.
\newblock {Cache-timing attacks on AES}.
\newblock 2005.

\bibitem{bhattacharyya2019smotherspectre}
A.~Bhattacharyya, A.~Sandulescu, et~al.
\newblock {Smotherspectre: Exploiting Speculative Execution Through Port Contention}.
\newblock {\em ACM SIGSAC Conference on Computer and Communications Security}, 2019.

\bibitem{bloem2022power}
R.~Bloem, B.~Gigerl, et~al.
\newblock {Power Contracts: Provably Complete Power Leakage Models for Processors}.
\newblock {\em ACM SIGSAC Conference on Computer and Communications Security}, pages 381--395, 2022.

\bibitem{chen2023psofuzz}
C.~Chen, V.~Gohil, et~al.
\newblock {PSOFuzz: Fuzzing Processors with Particle Swarm Optimization}.
\newblock {\em arXiv preprint arXiv:2307.14480}, 2023.

\bibitem{chen2022trusting}
C.~Chen, R.~Kande, et~al.
\newblock {Trusting the Trust Anchor: Towards Detecting Cross-Layer Vulnerabilities with Hardware Fuzzing}.
\newblock pages 1379--1383, 2022.

\bibitem{chen2023hypfuzz}
C.~Chen, R.~Kande, et~al.
\newblock {HyPFuzz: Formal-Assisted Processor Fuzzing}.
\newblock {\em USENIX Security Symposium}, pages 1361--1378, August 2023.

\bibitem{clarke2001progress}
E.~Clarke, O.~Grumberg, et~al.
\newblock {Progress on the State Explosion Problem in Model Checking}.
\newblock {\em Informatics}, pages 176--194, 2001.

\bibitem{clarke2018handbook}
E.~M. Clarke, T.~A. Henzinger, et~al.
\newblock {Handbook of Model Checking}.
\newblock 10, 2018.

\bibitem{clarke2011model}
E.~M. Clarke, W.~Klieber, et~al.
\newblock {Model Checking and the State Explosion Problem}.
\newblock {\em LASER Summer School on Software Engineering}, pages 1--30, 2011.

\bibitem{cyrluk1994effective}
D.~Cyrluk, S.~Rajan, et~al.
\newblock {Effective Theorem Proving for Hardware Verification}.
\newblock {\em International Conference on Theorem Provers in Circuit Design}, 1994.

\bibitem{dessouky2019hardfails}
G.~Dessouky, D.~Gens, et~al.
\newblock {HardFails: Insights into Software-Exploitable Hardware Bugs}.
\newblock {\em USENIX Security Symposium}, pages 213--230, 2019.

\bibitem{fadiheh2020formal}
M.~R. Fadiheh, J.~M{\"u}ller, et~al.
\newblock {A Formal Approach for Detecting Vulnerabilities to Transient Execution Attacks in Out-of-Order Processors}.
\newblock {\em ACM/IEEE Design Automation Conference}, pages 1--6, 2020.

\bibitem{fadiheh2019processor}
M.~R. Fadiheh, D.~Stoffel, et~al.
\newblock {Processor Hardware Security Vulnerabilities and Their Detection by Unique Program Execution Checking}.
\newblock {\em IEEE Design, Automation \& Test in Europe Conference \& Exhibition}, 2019.

\bibitem{ge2018survey}
Q.~Ge, Y.~Yarom, et~al.
\newblock {A Survey of Microarchitectural Timing Attacks and Countermeasures on Contemporary Hardware}.
\newblock {\em Journal of Cryptographic Engineering}, 2018.

\bibitem{gleissenthall2019iodine}
K.~v. Gleissenthall, R.~G. K{\i}c{\i}, et~al.
\newblock {IODINE: Verifying Constant-Time Execution of Hardware}.
\newblock {\em USENIX Security Symposium}, pages 1411--1428, 2019.

\bibitem{gohil2023mabfuzz}
V.~Gohil, R.~Kande, et~al.
\newblock {MABFuzz: Multi-Armed Bandit Algorithms for Fuzzing Processors}.
\newblock {\em arXiv preprint arXiv:2311.14594}, 2023.

\bibitem{gopinath2014code}
R.~Gopinath, C.~Jensen, et~al.
\newblock {Code Coverage for Suite Evaluation by Developers}.
\newblock {\em ACM/IEEE International Conference on Software Engineering}, pages 72--82, 2014.

\bibitem{gras2020absynthe}
B.~Gras, C.~Giuffrida, et~al.
\newblock {ABSynthe: Automatic Blackbox Side-channel Synthesis on Commodity Microarchitectures.}
\newblock {\em NDSS}, 2020.

\bibitem{IEEEstd}
S.~L.~W. Group.
\newblock {IEEE} {S}tandard for {S}ystem{V}erilog--{U}nified {H}ardware {D}esign, {S}pecification, and {V}erification {L}anguage.
\newblock {\em {IEEE} {S}td 1800-2017}, 2018.

\bibitem{gruss2016flush+}
D.~Gruss, C.~Maurice, et~al.
\newblock {Flush+ Flush: a fast and stealthy cache attack}.
\newblock {\em Detection of Intrusions and Malware, and Vulnerability Assessment}, pages 279--299, 2016.

\bibitem{guo2016scalable}
X.~Guo, R.~G. Dutta, et~al.
\newblock {Scalable SoC Trust Verification using Integrated Theorem Proving and Model Checking}.
\newblock pages 124--129, 2016.

\bibitem{hennessy2011computer}
J.~L. Hennessy and D.~A. Patterson.
\newblock {Computer Architecture: A Quantitative Approach}.
\newblock 2011.

\bibitem{hu2021hardware}
W.~Hu, A.~Ardeshiricham, et~al.
\newblock {Hardware Information Flow Tracking}.
\newblock {\em ACM Computing Surveys}, 2021.

\bibitem{hur2021difuzzrtl}
J.~Hur, S.~Song, et~al.
\newblock {DIFUZZRTL: Differential Fuzz Testing to Find CPU Bugs}.
\newblock {\em IEEE Symposium on Security and Privacy}, pages 1286--1303, 2021.

\bibitem{ibrahim2022microarchitectural}
A.~Ibrahim, H.~Nemati, et~al.
\newblock {Microarchitectural Leakage Templates and Their Application to Cache-Based Side Channels}.
\newblock {\em ACM SIGSAC Conference on Computer and Communications Security}, 2022.

\bibitem{ivankovic2019code}
M.~Ivankovi{\'c}, G.~Petrovi{\'c}, et~al.
\newblock {Code Coverage at Google}.
\newblock {\em ACM Joint Meeting on European Software Engineering Conference and Symposium on the Foundations of Software Engineering}, pages 955--963, 2019.

\bibitem{kandethehuzz}
R.~Kande, A.~Crump, et~al.
\newblock {TheHuzz: Instruction Fuzzing of Processors Using Golden-Reference Models for Finding Software-Exploitable Vulnerabilities}.
\newblock {\em USENIX Security Symposium}, pages 3219--3236, 2022.

\bibitem{Kocher2018spectre}
P.~Kocher, J.~Horn, et~al.
\newblock Spectre attacks: Exploiting speculative execution.
\newblock In {\em 40th IEEE Symposium on Security and Privacy (S\&P'19)}, 2019.

\bibitem{rfuzz}
K.~Laeufer, J.~Koenig, et~al.
\newblock {RFUZZ: Coverage-Directed Fuzz Testing of RTL on FPGAs}.
\newblock {\em IEEE International Conference on Computer-Aided Design}, 2018.

\bibitem{Lipp2018meltdown}
M.~Lipp, M.~Schwarz, et~al.
\newblock {Meltdown: Reading Kernel Memory from User Space}.
\newblock {\em {USENIX} Security}, 2018.

\bibitem{MITRE}
MITRE.
\newblock {CWE VIEW: Hardware Design}.
\newblock \url{https://cwe.mitre.org/data/definitions/1194.html}, 2019.
\newblock Last accessed on 09/28/2023.

\bibitem{muduli2020hyperfuzzing}
S.~K. Muduli, G.~Takhar, et~al.
\newblock {HyperFuzzing for SoC Security Validation}.
\newblock {\em ACM/IEEE International Conference on Computer-Aided Design}, pages 1--9, 2020.

\bibitem{riscvzkt}
M.-J. O.~Saarinen.
\newblock {riscv-zkt-list}.
\newblock \url{https://github.com/rvkrypto/riscv-zkt-list}, 2021.
\newblock Last accessed on 10/16/2023.

\bibitem{oleksenko2022revizor}
O.~Oleksenko, C.~Fetzer, et~al.
\newblock {Revizor: Testing Black-Box CPUs Against Speculation Contracts}.
\newblock {\em ACM International Conference on Architectural Support for Programming Languages and Operating Systems}, pages 226--239, 2022.

\bibitem{oleksenko2023hide}
O.~Oleksenko, M.~Guarnieri, et~al.
\newblock {Hide and Seek with Spectres: Efficient Discovery of Speculative Information Leaks with Random Testing}.
\newblock pages 1737--1752, 2023.

\bibitem{orenes2021autosva}
M.~Orenes-Vera, A.~Manocha, et~al.
\newblock {AutoSVA: Democratizing Formal Verification of RTL Module Interactions}.
\newblock {\em ACM/IEEE Design Automation Conference}, pages 535--540, 2021.

\bibitem{googlezenbleed}
T.~Ormandy and D.~Moghimi.
\newblock {Downfall and Zenbleed: Googlers helping secure the ecosystem}.
\newblock \url{https://security.googleblog.com/2023/08/downfall-and-zenbleed-googlers-helping.html}, 2023.
\newblock Last accessed on 09/28/2023.

\bibitem{osvik2006cache}
D.~A. Osvik, A.~Shamir, et~al.
\newblock {Cache Attacks and Countermeasures: The Case of AES}.
\newblock {\em The Cryptographers’ Track at the RSA Conference.}, 2006.

\bibitem{percival2005cache}
C.~Percival.
\newblock {Cache Missing for Fun and Profit}, 2005.

\bibitem{ragab2021crosstalk}
H.~Ragab, A.~Milburn, et~al.
\newblock {Crosstalk: Speculative Data Leaks Across Cores Are Real}.
\newblock {\em IEEE Symposium on Security and Privacy}, 2021.

\bibitem{rajapaksha2023sigfuzz}
C.~Rajapaksha, L.~Delshadtehrani, et~al.
\newblock {SIGFuzz: A Framework for Discovering Microarchitectural Timing Side Channels}.
\newblock In {\em 2023 Design, Automation \& Test in Europe Conference \& Exhibition (DATE)}, pages 1--6. IEEE, 2023.

\bibitem{riscv_home}
RISC-V.
\newblock {RISC-V Webpage}.
\newblock \url{https://riscv.org/}, 2023.
\newblock {Last accessed on 10/01/2023}.

\bibitem{robert2009monte}
C.~P. Robert.
\newblock {Monte Carlo Methods in Statistics}.
\newblock {\em arXiv:0909.0389}, 2009.

\bibitem{Schwarz2019ZombieLoad}
M.~Schwarz, M.~Lipp, et~al.
\newblock {ZombieLoad}: Cross-privilege-boundary data sampling.
\newblock {\em ACM SIGSAC Conference on Computer and Communications Security}, 2019.

\bibitem{shen2013modern}
J.~P. Shen and M.~H. Lipasti.
\newblock {Modern Processor Design: Fundamentals of Superscalar Processors}.
\newblock 2013.

\bibitem{verilator}
W.~Snyder.
\newblock {Verilator}.
\newblock \url{https://www.veripool.org/wiki/verilator}, 2023.
\newblock {Last accessed on 10/01/2023}.

\bibitem{solt2024cascade}
F.~Solt, K.~Ceesay-Seitz, et~al.
\newblock {Cascade: CPU Fuzzing via Intricate Program Generation}.
\newblock {\em USENIX Security Symposium}, 2024.

\bibitem{verifiwhitepaper}
Synopsys.
\newblock {Accelerating Verification Shift Left with Intelligent Coverage Optimization}.
\newblock \url{https://www.synopsys.com/cgi-bin/verification/dsdla/pdfr1.cgi?file=ico-wp.pdf}, 2022.
\newblock {Last accessed on 02/18/2023}.

\bibitem{trippel2018checkmate}
C.~Trippel, D.~Lustig, et~al.
\newblock {Checkmate: Automated Synthesis of Hardware Exploits and Security Litmus Tests}.
\newblock {\em IEEE International Symposium on Microarchitecture}, 2018.

\bibitem{tullsen1995simultaneous}
D.~M. Tullsen, S.~J. Eggers, et~al.
\newblock {Simultaneous Multithreading: Maximizing On-Chip Parallelism}.
\newblock {\em ACM International Symposium on Computer Architecture}, 1995.

\bibitem{van2018foreshadow}
J.~Van~Bulck, M.~Minkin, et~al.
\newblock {Foreshadow: Extracting the Keys to the Intel SGX Kingdom with Transient Out-of-Order Execution}.
\newblock {\em USENIX Security Symposium}, 2018.

\bibitem{vanbulck2020lvi}
J.~Van~Bulck, D.~Moghimi, et~al.
\newblock {LVI: Hijacking Transient Execution through Microarchitectural Load Value Injection}.
\newblock {\em IEEE Symposium on Security and Privacy}, 2020.

\bibitem{ridl}
S.~van Schaik, A.~Milburn, et~al.
\newblock {RIDL}: Rogue in-flight data load.
\newblock {\em IEEE Symposium on Security and Privacy}, 2019.

\bibitem{wang2014timing}
Y.~Wang, A.~Ferraiuolo, et~al.
\newblock {Timing Channel Protection for a Shared Memory Controller}.
\newblock 2014.

\bibitem{wang2023specification}
Z.~Wang, G.~Mohr, et~al.
\newblock {Specification and Verification of Side-Channel Security for Open-Source Processors via Leakage Contracts}.
\newblock {\em ACM SIGSAC Conference on Computer and Communications Security}, pages 2128--2142, 2023.

\bibitem{weaver2013non}
V.~M. Weaver, D.~Terpstra, et~al.
\newblock {Non-Determinism and Overcount on Modern Hardware Performance Counter Implementations}.
\newblock {\em IEEE International Symposium on Performance Analysis of Systems and Software}, pages 215--224, 2013.

\bibitem{weber2021osiris}
D.~Weber, A.~Ibrahim, et~al.
\newblock {Osiris: Automated Discovery of Microarchitectural Side Channels}.
\newblock {\em 30th USENIX Security Symposium}, pages 1--18, 2021.

\bibitem{wikner2022retbleed}
J.~Wikner and K.~Razavi.
\newblock {RETBLEED: Arbitrary speculative code execution with return instructions}.
\newblock {\em USENIX Security Symposium}, 2022.

\bibitem{wile2005comprehensive}
B.~Wile, J.~Goss, et~al.
\newblock {Comprehensive Functional Verification: The Complete Industry Cycle}.
\newblock 2005.

\bibitem{witharana2022survey}
H.~Witharana, Y.~Lyu, et~al.
\newblock {A Survey on Assertion-based Hardware Verification}.
\newblock {\em ACM Computing Surveys}, 2022.

\bibitem{xu2023morfuzz}
J.~Xu, Y.~Liu, et~al.
\newblock {MorFuzz: Fuzzing Processor via Runtime Instruction Morphing enhanced Synchronizable Co-simulation}.
\newblock 2023.

\bibitem{yavuz2022encider}
T.~Yavuz, F.~Fowze, et~al.
\newblock Encider: detecting timing and cache side channels in sgx enclaves and cryptographic apis.
\newblock {\em IEEE Transactions on Dependable and Secure Computing}, 20(2):1577--1595, 2022.

\bibitem{cva6}
F.~{Zaruba} and L.~{Benini}.
\newblock {The Cost of Application-Class Processing: Energy and Performance Analysis of a Linux-Ready 1.7-GHz 64-Bit RISC-V Core in 22-nm FDSOI Technology}.
\newblock {\em IEEE Transactions on Very Large Scale Integration Systems}, 2019.

\bibitem{boom}
J.~Zhao, B.~Korpan, et~al.
\newblock {SonicBOOM: The 3rd Generation Berkeley Out-of-Order Machine}.
\newblock {\em 4th Workshop on Computer Architecture Research with RISC-V}, 2020.

\bibitem{zhao2019send}
L.~Zhao, Y.~Duan, et~al.
\newblock {Send Hardest Problems My Way: Probabilistic path prioritization for hybrid fuzzing.}
\newblock {\em NDSS}, 2019.

\end{thebibliography}

\newpage
\appendix
\section*{Appendix} \label{apd:appendix}

\begin{table*}[]
\centering
\caption{Summary of results generated by \ourtool{} across different processors. Along with the detected vulnerability, we identify the specific lines in the RTL and trace the signals. We also note the time taken for detecting the various vulnerabilities.}
\resizebox{0.98\textwidth}{!}{
\begin{tabular}{|M{0.075\textwidth}|M{0.12\textwidth}|M{0.17\textwidth}|M{0.2\textwidth}|M{0.2\textwidth}|M{0.15\textwidth}| M{0.1\textwidth}|M{0.15\textwidth}|M{0.1\textwidth}|}
\hline
\textbf{Processor} & \textbf{Vulnerability} & \textbf{Source Module} & \textbf{RTL Lines} & \textbf{Phase 1 Results} & \textbf{Phase 2 Results} & \textbf{Seed generation(s)} & \textbf{Input generation ($*10^{3}$ s)} & \textbf{Leakage Analyzer ($*10^{4}$s)}\\
\hline
\cva{} & \texttt{DIVUW} & \texttt{serdiv} & 230115, 230117, 230166, 230169, 230173 & Multiple signals & Multiple Signals & 54.00 & 10.70 & 13.19\\
\hline
\cva{} & \texttt{Divide by zero} & \texttt{serdiv} & 13522, 13514, 13512 & Multiple signals & \texttt{div\_res\_zero\_q, op\_b\_q, state\_q, cnt\_q} & 3.67 & 0.61 & 2.00\\
\hline
\boom{} & \texttt{Divide by zero} & \texttt{MulDiv} & 230134, 230136, 230148, 230175 & Multiple Signals & \texttt{neg\_out, count, state, remainder} & 3.67 & 0.66 & 2.27\\
\hline
\rc{} & \texttt{Divide by zero} & \texttt{MulDiv} & 209671, 209673, 209703, 209725 & \texttt{io\_req\_bits\_in2, \_divisor\_T} & \texttt{neg\_out, divisor, state} & 3.67 & 0.63 & 1.65\\
\hline
\rc{} & \texttt{DIVUW} & \texttt{MulDiv} & 230115, 230117, 230166, 230173 & \texttt{io\_req\_bits\_in2, \_divisor\_T} & \texttt{\_divisor\_T,  divisor} & 3.67 & 0.53 & 1.71\\
\hline
\boom{} & \texttt{DIVUW} & \texttt{MulDiv} & 230115, 230117, 230166, 230173 & \texttt{io\_req\_bits\_in2, \_divisor\_T} & \texttt{\_divisor\_T, divisor} & 3.67 & 0.53 & 2.34\\
\hline
\boom{} & \texttt{SC} & \texttt{BoomWritebackUnit} & 180738, 180747, 180756, 180765, 180774, 180783, 180792, 180801 & \texttt{io\_data\_resp} & All \texttt{wb\_buffer} & 146.00 & 29.15 & 34.10\\
\hline
\cva{} & \texttt{Compressed Instructions} & \texttt{ALU} & 3757 & \texttt{adder\_z\_flag} & {Multiple Signals} & 124 & 24.77 & 1.77\\
\hline
\end{tabular}
}
\label{tab:DiagRes}
\end{table*}

\section{Implementation details of \graph{}}~\label{apd:imp_details}
\begin{figure}[!h]
    \centering
    \includegraphics[width=0.75\columnwidth,trim=0 2 6 0, clip]{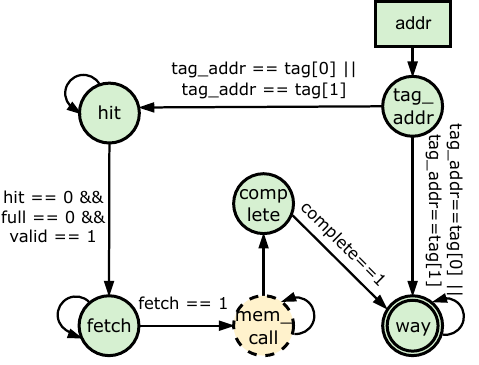}
    \caption{Sub-graph extracted from \graph{} in Listing~\ref{listing:case_study_full} of the multi-ways cache set protocol.}
    \label{fig:full_cs_graph}
\end{figure}

Section~\ref{sec:method_graph} explains the concept of a \graph{} (MEG). 
However, when implementing the strategy for real RTL designs, the syntaxes of hardware description languages~(HDLs), such as \verilog{} and \systemverilog{}~\cite{IEEEstd}, introduce more implementation challenges. Based on the example shown in Listing~\ref{listing:case_study}, we created a more complicated cache set with multiple ways as shown in Listing~\ref{listing:case_study_full}. 
We use the cache set as a case study to explain the implementation details of MEGs. Figure~\ref{fig:full_cs_graph} shows the sub-graph of the protocol.

\begin{lstlisting}[language=Verilog, label = {listing:case_study_full}, caption={Verilog code of multi-ways cache set protocol.},style=prettyverilog,float,belowskip=0pt,aboveskip=0pt,firstnumber=1]
input addr; output way;
reg hit, full, valid[cache_ways], fetch;
wire tag_addr, complete;
assign tag_addr = addr[tag_bits];
always@(posedge clock) begin
    if tag_addr == tag[0]
        way <= index[0], hit <= 1;
    else if tag_addr == tag[1]
        way <= index[1], hit <= 1;
    if hit == 0 && full != 1 begin
        if valid[0] == 1
             fetch <= 1; temp_way <= index;
        ...
    end
    if fetch == 1
        // Call mem_call for fetching block from next level.
        // Sub-instance sets complete signal
        mem_call(addr, complete);
    if complete:
        way <= temp_way
end
\end{lstlisting}

\textbf{Nested branch statements.} Except for the fundamental branch statements of HDLs, such as \texttt{if/else} and ternary (\texttt{?}), a register-transfer level (RTL) module also contains nested branch statements. 
Since all branch statements must be satisfied to trigger the operations under the most nested branch statement, we annotate an edge from each signal in those branch statements to the destination signals of the operations.
For example, from line 10 to 12, nested branch statements \texttt{hit $==$ 0 $\&\&$ full $\neq$ 1} and \texttt{valid[0] $==$ 1} drive the assignments of the \texttt{fetch} signal. 
Therefore, in Figure~\ref{fig:full_cs_graph}, the condition on the edge (\texttt{hit,fetch}) is the conjunction of \texttt{hit $==$ 0 $\&\&$ full $\neq$ 1} and \texttt{valid[0] $==$ 1}.

\textbf{Path Explosion:}
In certain cases, an event on one component is capable of triggering an event on another component under multiple different conditions. For example, from line 6 to 9, since the cache set has multiple ways, both condition \texttt{tag\_address == tag[0]} 
and \texttt{tag\_addr = tag[1]} can represent a cache hit.
This results in the existence of four edges from node \texttt{tag\_addr} to \texttt{hit} and \texttt{way}. Each is annotated with one condition between the two nodes. The existence of multiple edges between the same pair of nodes causes a path explosion during timing coverage instrumentation as mentioned in Section~\ref{sec:method_fuzz}. 
Consequently, the number of \systemverilog{} Assertion (SVA) \texttt{cover} properties in the module increases exponentially. 
To solve this challenge, we create a disjunction for different conditions on edges joining the same pair of nodes in a path.
Figure~\ref{fig:full_cs_graph} shows the disjunction on edge (\texttt{tag\_addr,hit}) and (\texttt{tag\_addr,way}).

\section{SVA \texttt{cover} properties for case study}~\label{apd:sva_case}

Listing~\ref{listing:sva_case} shows the \texttt{cover} property for the graphical paths in the \graph{} of the cache case study as shown in Listing~\ref{listing:case_study}. Section~\ref{sec:method_graph} mentions two \textit{Micro-Event Paths} from its input \texttt{addr} to its output \texttt{way}: \texttt{\{addr $\rightarrow$ tag\_addr $\rightarrow$ way\}} and \texttt{\{addr $\rightarrow$ tag\_addr $\rightarrow$ hit $\rightarrow$ fetch $\rightarrow$ mem\_call $\rightarrow$ complete $\rightarrow$ way\}}
Among these paths, \texttt{hit}, \texttt{full}, \texttt{fetch}, \texttt{way} are sequential nodes; the events occurring on these nodes complete on the next clock edge. The value of the \texttt{complete} signal is driven by \texttt{mem\_call}, a subinstance.

\begin{lstlisting}[language=Verilog, label = {listing:sva_case}, caption={SVA \texttt{cover} properties for case study.},style=prettyverilog,float=h,belowskip=0pt,aboveskip=0pt,firstnumber=1]
property p1; //{addr, tag_addr, way}
@(posedge clock) (tag_address == tag);
endproperty

property p2; //{addr, tag_addr, hit, fetch, mem_call, complete, way}
@(posedge clock) !(tag_address == tag) ##1 (hit == 0 && full != 1 && valid[0] == 1) |-> s_eventually (complete == 1);
endproperty
\end{lstlisting}\vspace{-0.4cm}

\section{Proof of concept codes for triggering detected vulnerabilities}~\label{apd:poc_trigger_code}
This section shows the proof of concept code snippets on triggering various timing vulnerabilities in different \riscv{} processors. 
Listing~\ref{listing:boom_divuw_rem_poc} shows the code snippets of \texttt{DIVUW + REM} side-channel on \boom{}~\cite{boom}.
Listing~\ref{listing:cva6_divuw_poc} shows the code snippets of \texttt{DIVUW} side-channel on \cva{}~\cite{cva6}.
Listing~\ref{listing:cva6_remw_poc} shows the code snippets of \texttt{REMW} side-channel on \cva{}~\cite{cva6}.
Listing~\ref{listing:cva6_other_poc} shows the code snippets of compressed instruction-based side-channels on \cva{}~\cite{cva6}.

\begin{lstlisting}[language=C, label = {listing:boom_divuw_rem_poc}, caption={\texttt{DIVUW + REM} side-channel proof of concept code snippets on \boom{}\cite{boom}.},style=customcArianeExploit,float=h,belowskip=0pt,aboveskip=0pt,firstnumber=1]
LI      a3, 291
LI      a7, -1954
LI      s10, 201
LI      t6, 477
// If t6=477, a7=-1954, s10=256, a3=291 takes 42ns
// If t6=1  , a7=-1954, s10=201, a3=291 takes 143ns
DIVUW   t5, a7, t6
REM     t3, s10, a3
\end{lstlisting}

\begin{lstlisting}[language=C, label = {listing:cva6_divuw_poc}, caption={\texttt{DIVUW} side-channel proof of concept code snippets on \cva{}\cite{cva6}.},style=customcArianeExploit,float=h,belowskip=0pt,aboveskip=0pt,firstnumber=1]
LI      a4, 3
// If a4=0 takes 92ns
// If a4=3 takes 36ns
LI      a7, 1333
// If a4=33 a7=-1333 takes 54ns
DIVUW   t3, a7, a4
\end{lstlisting}

\begin{lstlisting}[language=C, label = {listing:cva6_remw_poc}, caption={\texttt{REMW} side-channel proof of concept code snippets on \cva{}\cite{cva6}.},style=customcArianeExploit,float=h!,belowskip=0pt,aboveskip=0pt,firstnumber=1]
LI     a4, 3
// If a4=0 takes 92ns
// Else on average takes 37ns
LI     a7, 1333
REMW   t3, a7, a4
\end{lstlisting}

\begin{lstlisting}[language=C, label = {listing:cva6_other_poc}, caption={\texttt{C.ADD[W], C.SUB[W], C.AND, C.OR, C.XOR, and [C.]MV} side-channel proof of concept code snippets on \cva{}\cite{cva6}.},style=customcArianeExploit, float=h!,belowskip=0pt,aboveskip=0pt,firstnumber=1]
LI   a3, 0
// If a3=0 \& a4=0 takes 36 cycles
// Else on takes 24 cycles
LI   a4, 1023
MV   a3, a4
// Same for C.ADD[W], C.SUB[W], C.AND, C.OR, C.XOR, and C.MV
\end{lstlisting}

\section{Locations of side channels}~\label{apd:tv_location}
This section shows the locations of timing vulnerabilities identified by the \textit{Diagnozer} (See Section~\ref{sec:diagnozer}).
Listing~\ref{listing:div0_boom} shows the location of \textit{Division by zero} in \boom{}~\cite{boom}.
Listing~\ref{listing:div0_rc} shows the location of \textit{Division by zero} in \rc{}~\cite{rocket_chip_generator}.
Listing~\ref{Listing:div0_cva} shows the location of \textit{Division by zero} in \cva{}~\cite{cva6}.
Listing~\ref{listing:div_rem} shows the location of \texttt{DIVUW+REM} in \boom{}.
Listing~\ref{listing:ci_cva} shows the location of \textbf{compressed instructions} and \textbf{\texttt{MV}} in \cva{}; nine vulnerabilities share the same root cause.
The results show that the \textit{Diagnozer} of \ourtool{} can successfully identify the location of timing vulnerabilities in processors. 

        \bgroup
        \createlinenumber{1}{230133}
        \createlinenumber{2}{230134}
        \createlinenumber{3}{230135}
        \createlinenumber{4}{230136}
        \createlinenumber{5}{230137}
        \createlinenumber{6}{230138}
        \createlinenumber{7}{230139}
        \createlinenumber{8}{230140}
        \createlinenumber{9}{$\dots$}
        \createlinenumber{10}{230147}
        \createlinenumber{11}{230148}
        \createlinenumber{12}{230149}
        \createlinenumber{13}{230150}
        \createlinenumber{14}{230151}
        \createlinenumber{15}{230152}
        \createlinenumber{16}{$\dots$}
        \createlinenumber{17}{230174}
        \createlinenumber{18}{230175}
        \createlinenumber{19}{230176}
\begin{lstlisting}[language=Verilog, label = {listing:div0_boom}, caption={Source location of \textbf{\textit{Division by zero}} in \boom{}~\cite{boom}.},style=prettyverilog,float=h!,belowskip=0pt,aboveskip=0pt,firstnumber=1, escapechar=\%,linewidth=\linewidth, xleftmargin=25pt]
if (eOut_1 == 1) begin // @[Multiplier.scala 153:19]
    %\Hilight[0.48]%count <= {{1'd0}, eOutPos}; 
end else begin
    %\Hilight[0.35]%count <= _count_T_1; // @[Multiplier.scala 143:11]
end
...
if (divby0 & _eOut_T_4) begin
    %\Hilight[0.29]%neg_out <= 1'h0; // @[Multiplier.scala 158:38]
end
...
end else if (state == 3'h3) begin
    %\Hilight[0.54]%remainder <= {{1'd0}, _GEN_16};
end
\end{lstlisting}
        \egroup

        \bgroup
        \createlinenumber{1}{209670}
        \createlinenumber{2}{209671}
        \createlinenumber{3}{209672}
        \createlinenumber{4}{209673}
        \createlinenumber{5}{209674}
        \createlinenumber{6}{$\dots$}
        \createlinenumber{7}{209702}
        \createlinenumber{8}{209703}
        \createlinenumber{9}{$\dots$}
        \createlinenumber{10}{209724}
        \createlinenumber{11}{209725}
\begin{lstlisting}[language=Verilog, label = {listing:div0_rc}, caption={Source location of \textbf{\textit{Division by zero}} in \rc{}~\cite{rocket_chip_generator}.},style=prettyverilog,float=h,belowskip=0pt,aboveskip=0pt,firstnumber=1, escapechar=\%,linewidth=\linewidth, xleftmargin=25pt]
end else if (lhs_sign | rhs_sign) begin // @[Multiplier.scala 164:36]
    %\Hilight[0.25]%state <= 3'h1;
end else begin
    %\Hilight[0.25]%state <= 3'h3;
end
...
end else begin
    %\Hilight[0.56]%neg_out <= lhs_sign != rhs_sign;
...
if ( _T_38) begin // @[Multiplier.scala 163:24]
    %\Hilight[0.40]%divisor <= _divisor_T; // @[Multiplier.scala 169:13]
\end{lstlisting}
        \egroup

        \bgroup
        \createlinenumber{1}{13512}
        \createlinenumber{2}{13513}
        \createlinenumber{3}{13514}
        \createlinenumber{4}{13515}
        \createlinenumber{5}{13516}
        \createlinenumber{6}{13517}
        \createlinenumber{7}{13518}
        \createlinenumber{8}{13519}
        \createlinenumber{9}{13520}
        \createlinenumber{10}{13521}
        \createlinenumber{11}{13522}
\begin{lstlisting}[language=Verilog, label = {Listing:div0_cva}, caption={Source location of \textbf{\textit{Division by zero}} in \cva{}~\cite{cva6}.},style=prettyverilog,float=h,belowskip=0pt,aboveskip=0pt,firstnumber=1, escapechar=\%,linewidth=\linewidth, xleftmargin=25pt]
    %\Hilight[0.34]%state_q        <= state_d;
    op_a_q         <= op_a_d;
    %\Hilight[0.30]%op_b_q         <= op_b_d;
    res_q          <= res_d;
    %\Hilight[0.26]%cnt_q          <= cnt_d;
    id_q           <= id_d;
    rem_sel_q      <= rem_sel_d;
    comp_inv_q     <= comp_inv_d;
    res_inv_q      <= res_inv_d;
    op_b_zero_q    <= op_b_zero_d;
    %\Hilight[0.57]%div_res_zero_q <= div_res_zero_d;
\end{lstlisting}
        \egroup

        \bgroup
        \createlinenumber{1}{230112}
        \createlinenumber{2}{230113}
        \createlinenumber{3}{230114}
        \createlinenumber{4}{230115}
        \createlinenumber{5}{230116}
        \createlinenumber{6}{230117}
        \createlinenumber{7}{$\dots$}
        \createlinenumber{8}{230165}
        \createlinenumber{9}{230166}
        \createlinenumber{10}{230167}
        \createlinenumber{11}{230168}
        \createlinenumber{12}{230169}
        \createlinenumber{13}{230170}
        \createlinenumber{14}{230171}
        \createlinenumber{15}{230172}
        \createlinenumber{16}{230173}
\begin{lstlisting}[language=Verilog, label = {listing:div_rem}, caption={Source location of \textbf{\texttt{DIVUW+REM}} in \boom{}~\cite{boom}.},style=prettyverilog,float=h,belowskip=0pt,aboveskip=0pt,firstnumber=1, escapechar=\%, linewidth=\linewidth, xleftmargin=25pt]
if (cmdMul) begin // @[Multiplier.scala 164:17]
    state <= 3'h2;
end else if (lhs_sign | rhs_sign) begin // @[Multiplier.scala 164:36]
    %\Hilight[0.25]%state <= 3'h1;
end else begin
    %\Hilight[0.25]%state <= 3'h3;
...
if ( _T_38) begin // @[Multiplier.scala 163:24]
    %\Hilight[0.40]%divisor <= _divisor_T; // @[Multiplier.scala 169:13]
end else if (state == 3'h1) begin // @[Multiplier.scala 91:57]
    if (divisor[63]) begin // @[Multiplier.scala 95:25]
        divisor <= subtractor; // @[Multiplier.scala 96:15]
    end
end
if ( _T_38) begin // @[Multiplier.scala 163:24]
    %\Hilight[0.54]%remainder <= {{66'd0}, lhs_in}; // @[Multiplier.scala 170:15]
\end{lstlisting}
        \egroup

        \bgroup
        
        \createlinenumber{1}{3754}
        \createlinenumber{2}{3755}
        \createlinenumber{3}{3756}
        \createlinenumber{4}{3757}
\begin{lstlisting}[language=Verilog, label = {listing:ci_cva}, caption={Source location of \textbf{compressed instructions} and \textbf{\texttt{MV}} in \cva{}~\cite{cva6}.},style=prettyverilog,float=h,belowskip=0pt,aboveskip=0pt,firstnumber=1, escapechar=\%,linewidth=\linewidth, xleftmargin=25pt]
assign adder_z_flag       = ~|adder_result;

// get the right branch comparison result
always_comb begin : branch_resolve
    // set comparison by default
    alu_branch_res_o      = 1'b1;
    case (fu_data_i.operator)
        %\Hilight[0.62]%EQ:       alu_branch_res_o = adder_z_flag;
\end{lstlisting}
        \egroup
        
\begin{table}[]
    \centering
    \caption{MEG and SVA overhead statistics for \boom{}~\cite{boom}.}
    \label{tab:overhead_stats}
    \begin{tabular}{|M{0.28\columnwidth}|M{0.15\columnwidth}|M{0.2\columnwidth}|M{0.15\columnwidth}|}
\hline
        Module & \multicolumn{2}{c|}{MEG} & SVA coverage
        points\\
        \cline{2-3}
        & Time taken(s) & Space consumed(kB) & (per module)\\
        \hline
        MaxPFibLFSR1 & 0.32 & 5.2 & 302\\
        \hline
        BoomIOMSHR & 0.34 & 23.1 & 157\\
        \hline
        BoomWbUnit & 0.32 & 20.6 & 3372 \\
        \hline
        FetchBuffer & 0.52 & 403.2 & 1458 \\
        \hline
        BrnchKillQ2 & 0.35 & 22.7 & 1066 \\
        \hline
        ICache & 0.13 & 14.9 & 3869 \\
        \hline
        TLXbar\_7 & 0.32 & 13.1 & 187 \\
        \hline
        BrnchKillQ & 0.49 & 106.1 & 2777\\
        \hline
        BoomRAS & 0.35 & 17.0 & 125 \\
        \hline
        Queue\_35 & 0.36 & 20.7 & 124 \\
        \hline
    \end{tabular}
\end{table}
\end{document}